\documentclass[colorlinks,bookmarks=false,citecolor=blue,linkcolor=red,urlcolor=blue]{SciPost}
\usepackage{hyperref}
\usepackage{graphicx}

\usepackage[square,numbers,sort&compress]{natbib}

\begin{document}

\begin{center}{\Large \textbf{
Efficient Quantum Monte Carlo simulations of highly frustrated magnets:
the frustrated spin-1/2 ladder
}}\end{center}

\begin{center}
S.~Wessel,\textsuperscript{1} 
B.~Normand,\textsuperscript{2}
F.~Mila,\textsuperscript{3}
A.~Honecker\textsuperscript{4*}
\end{center}

\begin{center}
{\bf 1} Institut f\"ur Theoretische Festk\"orperphysik, JARA-FIT and 
JARA-HPC, \\ RWTH Aachen University, D-52056 Aachen, Germany
\\
{\bf 2}
Laboratory for Neutron Scattering and Imaging,
Paul Scherrer Institute, \\ CH-5232 Villigen-PSI, Switzerland
\\
{\bf 3} Institute of Physics, Ecole Polytechnique F\'ed\'erale
Lausanne (EPFL), \\ CH-1015 Lausanne, Switzerland
\\
{\bf 4} Laboratoire de Physique Th\'eorique et Mod\'elisation, CNRS 
UMR 8089, \\ Universit\'e de Cergy-Pontoise, 
F-95302 Cergy-Pontoise Cedex, France
\\
* andreas.honecker@u-cergy.fr
\end{center}

\date{\today}
\begin{abstract}

Quantum Monte Carlo simulations provide one of the more powerful and 
versatile numerical approaches to condensed matter systems. However, their 
application to frustrated quantum spin models, in all relevant temperature 
regimes, is hamstrung by the infamous ``sign problem.'' Here we exploit the 
fact that the sign problem is basis-dependent. Recent studies have shown that 
passing to a dimer (two-site) basis eliminates the sign problem completely 
for a fully frustrated spin model on the two-leg ladder. We generalize this 
result to all partially frustrated two-leg spin-1/2 ladders, meaning those 
where the diagonal and leg couplings take any antiferromagnetic values. We 
find that, although the sign problem does reappear, it remains remarkably 
mild throughout the entire phase diagram. We explain this result and apply 
it to perform efficient quantum Monte Carlo simulations of frustrated 
ladders, obtaining accurate results for thermodynamic quantities such 
as the magnetic specific heat and susceptibility of ladders up to $L = 200$ 
rungs (400 spins 1/2) and down to very low temperatures.

\end{abstract}

\vspace{10pt}
\noindent\rule{\textwidth}{1pt}
\tableofcontents\thispagestyle{fancy}
\noindent\rule{\textwidth}{1pt}
\vspace{10pt}

\section{Introduction}

\label{sec:Intro}

Quantum Monte Carlo (QMC) simulation ranks among the most general and 
efficient methods for studying both the static and dynamic properties of 
quantum magnets at finite temperatures \cite{EvertzLoop03,SandvikComp10}. 
An early landmark example was provided by the square-lattice antiferromagnet, 
where field-theoretical predictions \cite{PhysRevB.39.2344,Hasenfratz1993,
PhysRevB.49.11919} were tested by QMC simulations for the spin-1/2 
\cite{PhysRevLett.80.2705} and spin-1 cases \cite{HTN98}. A second 
valuable type of system was the $n$-leg spin ladder, which emerged as a 
tool for understanding the two-dimensional cuprates from the limit of one 
dimension \cite{DR96,Dagotto99,Mikeska2004}, and where QMC simulations 
were essential for understanding the spin gap, correlation length, and 
magnetic susceptibility \cite{Hida91,PhysRevB.47.3196,PhysRevLett.77.1865,
PhysRevB.54.R3714,ladderQMCrev}.

Geometrically frustrated magnets constitute an important class of quantum 
spin system with the potential to host exotic phases such as the quantum 
spin liquid \cite{Richter2004,Balents10,HFMbook,DIEPbook,Liao17}. However, 
QMC simulations on geometrically frustrated lattices suffer from the notorious 
``sign problem,'' which is the appearance of spin configurations with negative 
weights; a detailed discussion is deferred to Sec.~\ref{sec:QMC}. This 
problem restricts conventional QMC simulations to systems that are at most 
weakly frustrated \cite{Miyahara98,ladderQMCrev,PhysRevB.79.104432,
PhysRevB.84.094445}. Although a general solution to the sign problem is not 
to be expected \cite{PhysRevLett.94.170201}, progress has nevertheless been 
possible in some cases \cite{PhysRevB.57.R3197,PhysRevB.58.2411,
PhysRevLett.83.3116,PhysRevB.62.1102,PhysRevB.70.100410,PhysRevLett.100.247206,
PhysRevB.89.134422,Hann201763}. Specifically, for certain highly frustrated 
magnets, the Hamiltonian may be reexpressed in terms of cluster spins; if 
these form a bipartite lattice, sign-free QMC simulations are possible in 
the cluster basis \cite{PhysRevB.93.054408,PhysRevLett.117.197203,NgYang2017}. 
For the example of a frustrated ladder, the clusters correspond to the ladder 
rungs. In the present manuscript, we will go away from the case of perfect 
frustration, where the sign problem can be eliminated completely 
\cite{PhysRevB.93.054408,PhysRevLett.117.197203}. We will show that, 
although a sign problem remains present, it is so mild that the cluster 
basis allows efficient QMC simulations at all points in the phase diagram 
of the frustrated antiferromagnetic spin-1/2 ladder.

The structure of this article is as follows. We begin in Sec.~\ref{sec:Model} 
by presenting the model in detail. The QMC methods that we use to compute 
thermodynamic properties are introduced in Sec.~\ref{sec:QMC}, where the sign 
problem and its manifestations in the cluster basis are discussed in detail. 
Section \ref{sec:Results} presents QMC results for the magnetic specific heat 
and susceptibility for sets of parameter choices representative of every 
region of the phase diagram (shown by the black dots in 
Fig.~\ref{fig:ladder_phaseDiag}), which we compare with a range of numerical 
and theoretical results as an aid to physical interpretation. We summarize 
and offer some perspectives for reduced-sign-problem QMC in Sec.~\ref{sec:Sum}.

\section{Model: Frustrated ladder}

\label{sec:Model}

\subsection{Hamiltonian and conservation laws}

\begin{figure}[t!]
\centering\includegraphics[width=0.8\columnwidth]{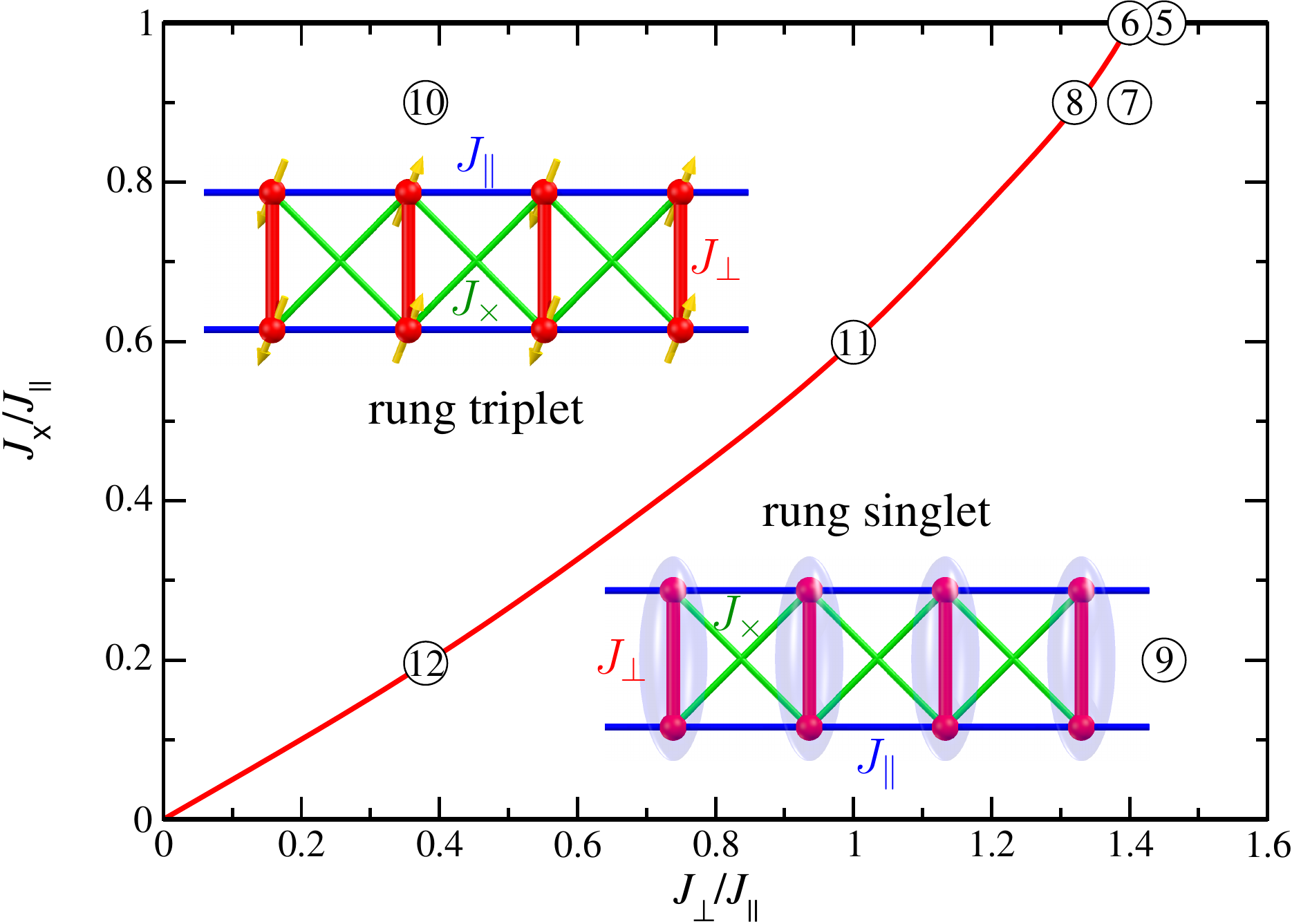}
\caption{Ground-state phase diagram of the frustrated spin ladder, as 
established by a range of numerical methods \cite{PhysRevB.57.11439,Wang2000,
PhysRevB.73.224433,PhysRevB.77.205121,PhysRevB.77.214418,PhysRevB.86.075133,
Chen2016}. The system shows only two phases, rung-singlet and rung-triplet, 
which are separated by a quantum phase transition (red line). In the inset 
schematics, the ladder sites (spheres) host $S = 1/2$ quantum spins and the 
Heisenberg couplings between spins are specified by the parameters $J_\perp$ 
for the ladder rungs, $J_\|$ for the ladder legs (blue), and $J_\times$ for 
the cross-plaquette couplings (green), which we take to be symmetrical. 
Purple rungs with ellipses represent rung-singlet spin states and red 
rungs with two parallel spins represent rung triplets. The numbered 
circles designate those points in the phase diagram for which we present 
thermodynamic results, each number matching that of the corresponding 
figure in Sec.~\ref{sec:Results}.}
\label{fig:ladder_phaseDiag}
\end{figure}

The Hamiltonian of a frustrated two-leg ladder with $L$ rungs, for any spin 
quantum number, $S$, is 
\begin{equation} 
H =  J_\perp \sum_{i} {\vec S}_{i}^1 \cdot {\vec S}_{i}^2
+ J_\| \sum_{i,m=1,2}  {\vec S}_{i}^m \cdot {\vec S}_{i+1}^m
+ J_\times \sum_{i,m=1,2} {\vec S}_{i}^m \cdot {\vec S}_{i+1}^{\bar m} ,
\label{eq:essh} 
\end{equation} 
where $i$ is the rung index, $m = 1$ and $2$ denote the two chains of the 
ladder, and ${\bar m}$ is the chain opposite to $m$. The superexchange 
parameters, $J_\perp$, $J_\|$, and $J_\times$ are depicted in the insets of 
Fig.~\ref{fig:ladder_phaseDiag} and we comment that the ladder we consider 
is always symmetrical under reflection through an axis bisecting all its 
rungs (i.e.~under exchange of chains 1 and 2). In our numerical calculations 
we will impose periodic boundary conditions, such that $i + L \equiv i$.

Let us introduce the total-spin and spin-difference operators on rung $i$, 
\begin{equation}
\vec{T}_i = \vec{S}_{i}^1 + {\vec S}_{i}^2 \, , \qquad
\vec{D}_i = \vec{S}_{i}^1 - {\vec S}_{i}^2 \, .
\label{eq:defTD} 
\end{equation}
The $SU(2)$ algebra of the operators $\vec{S}_{i}^m$ implies the 
on-site commutation relations 
\begin{equation}
\left[T_i^\alpha,T_i^\beta\right] = \!\! \sum_{\gamma=x,y,z} \!\!  
\epsilon^{\alpha,\beta}_{\phantom{\alpha,\beta}\gamma} \, T_i^\gamma \, , \quad
\left[T_i^\alpha,D_i^\beta\right] = \!\! \sum_{\gamma=x,y,z} \!\! 
\epsilon^{\alpha,\beta}_{\phantom{\alpha,\beta}\gamma} \, D_i^\gamma \, , \quad
\left[D_i^\alpha,D_i^\beta\right] = \!\! \sum_{\gamma=x,y,z} \!\! 
\epsilon^{\alpha,\beta}_{\phantom{\alpha,\beta}\gamma} \, T_i^\gamma
\label{eq:TDcomm} 
\end{equation}
for the operators of Eq.~(\ref{eq:defTD}), where $\alpha,\beta = x,y,z$ are 
the Cartesian components of the spin operators and the commutators for 
different sites vanish automatically.

Using the composite operators (\ref{eq:defTD}), the Hamiltonian (\ref{eq:essh}) 
can be reexpressed in the form
\begin{equation} 
H = \sum_{i=1}^L \left( J_\perp \, \left(\frac{1}{2} \, \vec{T}_i^2
 - S\,(S+1)\right)
+ \frac{J_\| + J_\times}{2} \, \vec{T}_i \cdot \vec{T}_{i+1} 
+ \frac{J_\| - J_\times}{2} \, \vec{D}_i \cdot \vec{D}_{i+1} \right) .
\label{eq:exehGen} 
\end{equation}
We note that the exchange of leg and diagonal couplings, $J_\|$ and $J_\times$, 
yields an equivalent Hamiltonian: the first term of Eq.~(\ref{eq:exehGen})
is manifestly invariant and the second symmetric under the exchange of $J_\|$ 
and $J_\times$, while the last changes sign. However, this sign-change is easily 
compensated by exchanging the order of legs $1$ and $2$ on every second rung 
in the transformation (\ref{eq:defTD}). It follows that the system is 
symmetric under the interchange of $J_\|$ and $J_\times$ and therefore we 
restrict our considerations to the regime $J_\times \le J_\|$.

At $J_\times = J_\|$, the final term in Eq.~(\ref{eq:exehGen}) disappears
and the expression simplifies to \cite{PhysRevB.52.12485,Honecker2000}
\begin{equation} 
H = J_\| \sum_{i=1}^L \vec{T}_i \cdot \vec{T}_{i+1} + J_\perp \sum_{i=1}^L 
\left( \frac{1}{2} \, \vec{T}_i^2 - S\,(S + 1) \right) .
\label{eq:exeh} 
\end{equation}
At this fully frustrated point, the Hamiltonian (\ref{eq:essh}) has $L$ 
purely local conservation laws, namely the total spin $\vec{T}_i^2$ on each 
individual rung, which may be encoded in additional quantum numbers, $T_i$.
Although the form of Eq.~(\ref{eq:exeh}) is valid for all $S$ 
\cite{PhysRevB.92.115111}, henceforth we consider exclusively the case 
$S = 1/2$, where $T_i$ takes the values $0$ (a rung singlet) or $1$ (a 
rung triplet).

\subsection{Ground-state phase diagram}


Before discussing the finite-temperature properties of the frustrated spin-1/2 
ladder, it is useful to recall its ground-state phase diagram, which is shown 
in Fig.~\ref{fig:ladder_phaseDiag}. Considering first the unfrustrated case, 
$J_\times = 0$, in the limit of strong rung coupling, $J_\perp \gg J_\|$, the 
system clearly adopts a gapped rung-singlet state \cite{PhysRevB.47.3196,
RTR94}. By contrast, for $J_\perp = 0 = J_\times$, one has decoupled spin-1/2 
chains, which are known to be gapless \cite{PhysRevLett.45.1358,FADDEEV1981,
Mikeska2004}. While early numerical work suggested the possibility that a 
finite critical value of $J_\perp$ may be required to open the gap 
\cite{Hida91,PhysRevB.45.5744}, scaling and field-theory arguments led to the 
conclusions that the critical value vanishes and the gap scales linearly with 
$J_\perp > 0$ \cite{PhysRevLett.69.2419,HatanoNishiyama95,PhysRevB.53.8521}. 
These results were confirmed by later numerical work \cite{PhysRevB.47.3196,
NHS95,PhysRevLett.77.1865} and the critical value $J_{\perp,c} = 0$ is now well 
established for $J_\times = 0$.

The other well-controlled case is the fully frustrated situation, $J_\times
 = J_\|$. The observations of the previous subsection explain the phase 
diagram along this line: the last term in Eq.~(\ref{eq:exeh}) enforces all 
$T_i = 0$ for large $J_\perp$ (rung-singlet phase) while for small $J_\perp$ 
the fluctuations of the first term dominate, giving all $T_i = 1$ 
(rung-triplet phase) \cite{PhysRevB.43.8644,PhysRevB.48.10653,
PhysRevB.52.12485,Honecker2000}. It is firmly established that this is 
a direct first-order transition, and takes place at a critical coupling 
$J_{\perp,c} \simeq 1.401484\,J_\|$ \cite{PhysRevB.43.8644,PhysRevB.48.10653,
PhysRevB.52.12485,Honecker2000}, which is inferred from numerical results 
for the ground-state energy of the spin-1 chain \cite{PhysRevB.48.3844,
PhysRevB.50.3037}. The first-order phase transition remains present for small 
deviations away from perfect frustration, $J_\times \neq J_\|$, and as such is 
easy to trace numerically in this regime \cite{PhysRevB.57.11439,Wang2000}.

Returning to the limit of weakly coupled chains, $J_\perp,J_\times \ll J_\|$, 
a field-theoretical analysis \cite{PhysRevB.61.8871,PhysRevLett.93.127202} 
predicted that the line separating these two phases approaches $J_\perp = 
2\,J_\times$. In fact the field theory further predicts an intermediate 
columnar-dimer phase \cite{PhysRevLett.93.127202}, although such a phase 
has not been observed unambiguously in any numerical investigations of 
the antiferromagnetic frustrated ladder \cite{PhysRevB.57.11439,Wang2000,
PhysRevLett.105.077202} despite repeated efforts to verify its existence 
\cite{PhysRevB.73.224433,PhysRevB.77.205121,PhysRevB.77.214418,
PhysRevB.86.075133,Chen2016}. By contrast, an intermediate columnar-dimer 
phase can be stabilized by ferromagnetic superexchange couplings 
\cite{PhysRevB.81.064432} or by an additional next-nearest-neighbor coupling, 
$J_2$, along the ladder legs \cite{PhysRevB.73.214427,PhysRevB.77.214418,
RGHS11,LiLin12}. Although the frustrated ladder has to date been studied 
primarily from a theoretical perspective, the recently synthesized compound 
Li$_2$Cu$_2$O(SO$_4$)$_2$ \cite{ACS.chemmater.5b00588,RRGSVR17} has a ladder 
geometry with $J_\times = J_\|$, although in this case $J_\perp$ is ferromagnetic 
and there is a significant antiferromagnetic $J_2$ \cite{VRSR17}.

In the following we will focus on the antiferromagnetic case with $J_2 = 0$. 
Figure \ref{fig:ladder_phaseDiag} summarizes numerical results that have been 
obtained for the full phase diagram of the antiferromagnetic spin-1/2 ladder 
\cite{PhysRevB.57.11439,Wang2000,PhysRevB.73.224433,PhysRevB.77.205121, 
PhysRevB.77.214418,PhysRevB.86.075133,Chen2016}. These interpolate the 
rung-triplet-to-rung-singlet quantum phase transition between the analytically 
known limits of a direct, first-order process at large $J_\perp$ to a complex 
process with the possibility of an invisibly narrow intermediate phase at 
small $J_\perp$.

\section{Method: Quantum Monte Carlo simulations}

\label{sec:QMC}

\subsection{Sign problem}

Our aim is to use QMC simulations to compute thermodynamic quantities 
for the frustrated ladder. The standard approach for dealing with negative 
weights appearing for some configurations is to use a reweighting scheme 
that performs the QMC sampling with respect to their absolute values. 
However, this requires keeping track of the sign of each configuration 
and including it in any measurement, as discussed in Refs.~\cite{EvertzLoop03,
PhysRevLett.94.170201}. The performance of the simulation is then determined 
by the average sign, $\langle \text{sign}\rangle$: as long as this quantity 
is close to unity, Monte Carlo sampling is still efficient, whereas a small 
average sign must be compensated by a corresponding increase in the number 
of samples. Stated precisely, Monte Carlo errors decrease with the square 
root of the number of samples, and thus to compensate, for example, for an 
average sign of $10^{-2}$ it is necessary to run the code $10^4$ times longer.

\begin{figure}[t!]
\centering\includegraphics[width=0.66\columnwidth]{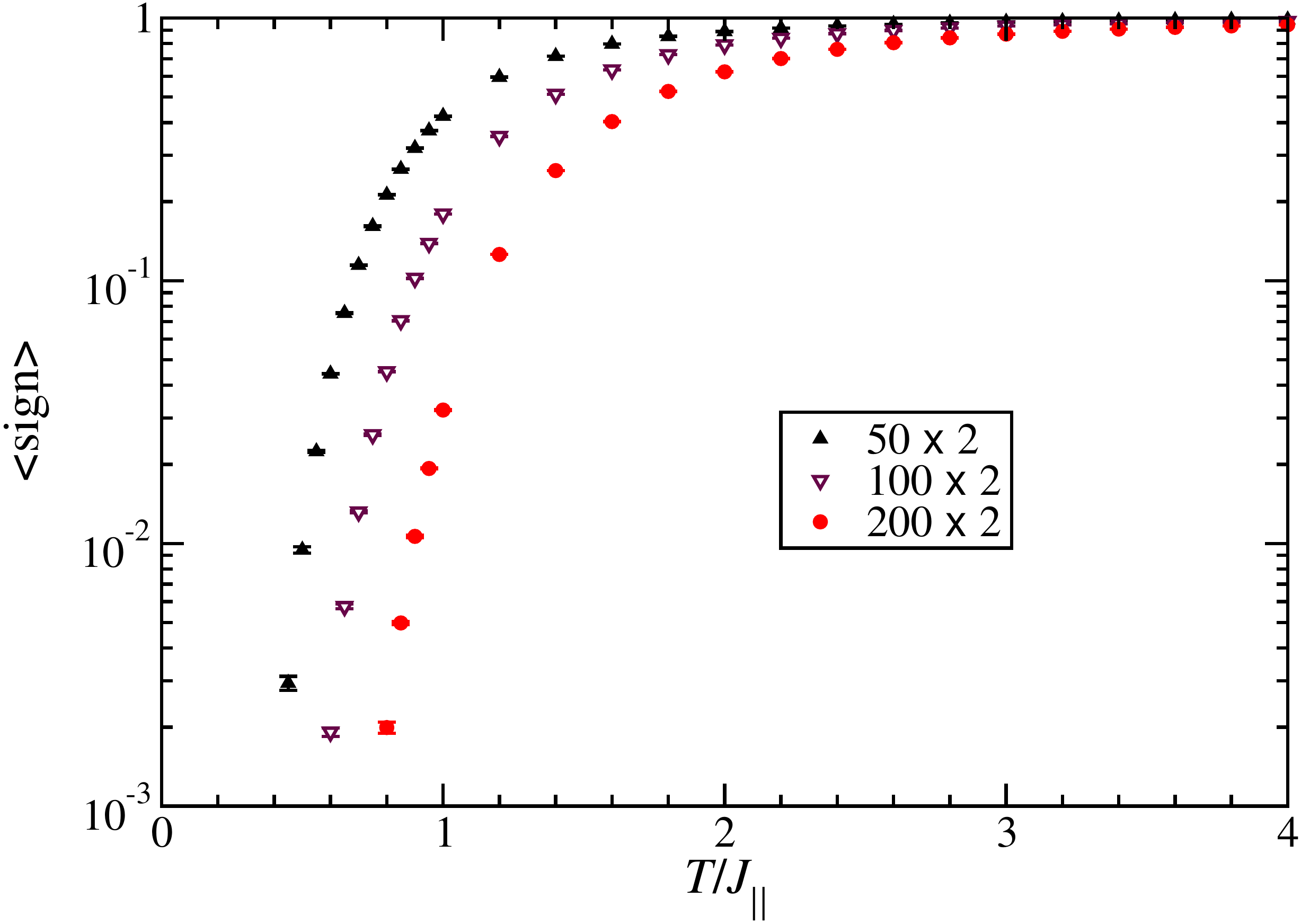}
\caption{Average sign, $\langle \text{sign} \rangle$, for a QMC simulation
in the single-site basis of the frustrated ladder at $J_\perp = 0.38$, 
$J_\| = 1$, and $J_\times = 0.196$, shown as a function of temperature for 
ladders with $L = 50$, $100$, and $200$ rungs. Analogous results for the 
rung basis are shown in Fig.~\ref{fig:signJR0_38JD0_196} and the 
thermodynamic response, also computed in the rung basis, in 
Fig.~\ref{fig:thermoBreakJR0_38JD0_196}.}
\label{fig:signJR0_38JD0_196site}
\end{figure}

Throughout most of the phase diagram of the frustrated ladder, QMC simulations 
of the Hamiltonian in the single-site basis (\ref{eq:essh}) suffer from a 
severe sign problem. This was illustrated for a case on the fully frustrated 
line ($J_\times = J_\|$) in Fig.~8 of Ref.~\cite{PhysRevB.93.054408}. Here we 
present in Fig.~\ref{fig:signJR0_38JD0_196site} an illustration of the 
average sign for a less strongly frustrated case. We note that the scale of 
$\langle \text{sign} \rangle$ is logarithmic, and thus that there remains an 
exponential suppression both with decreasing temperature and with increasing 
system size \cite{PhysRevLett.94.170201}. For a system with $L = 200$ rungs, 
the average sign drops below $10^{-2}$ for $T \lesssim J_\|$, rendering the 
low-temperature region completely inaccessible for systems of any meaningful 
size.

While the structure and nature of the sign problem allow little hope that it 
can be overcome in general \cite{PhysRevLett.94.170201}, its severity does 
still depend on the choice of basis. As noted in Sec.~\ref{sec:Intro}, 
frustrated quantum spin models have been identified where the sign problem 
can indeed be overcome, and their number continues to grow. Leading methods 
for tackling the sign problem to date include meron and nested cluster 
algorithms \cite{PhysRevLett.83.3116,PhysRevB.62.1102,PhysRevLett.100.247206}
and a suitable choice of simulation basis \cite{PhysRevB.57.R3197,
PhysRevB.58.2411,PhysRevB.70.100410,PhysRevB.89.134422,PhysRevB.93.054408,
PhysRevLett.117.197203,Hann201763,NgYang2017,PhysRevB.92.195126}. For the 
frustrated ladder, we follow the latter approach, taking the rung basis as 
a natural choice for the rewritten Hamiltonian (\ref{eq:exehGen}).

\subsection{Rung basis}

\subsubsection{Fully frustrated ladder}

At the fully frustrated point, $J_\times = J_\|$, it is clear from the 
vanishing of the $\vec{D} \cdot \vec{D}$ term in Eq.~(\ref{eq:exehGen}) 
that the sign problem is eliminated completely [Eq.~(\ref{eq:exeh})] 
\cite{PhysRevB.93.054408,PhysRevLett.117.197203}. We choose to sample 
the partition function in this basis using the stochastic series 
expansion (SSE) representation with generalized directed loop updates
\cite{PhysRevE.66.046701,PhysRevE.71.036706}. However, the fact that the 
Hilbert space is split into sectors characterized by the different local
quantum numbers $\{T_i\}$ gives rise to problems with the ergodicity of
the conventional SSE algorithm at low temperatures, and these require a 
parallel-tempering protocol \cite{IJMPC96,JPSJ96,Marinari1998} to overcome. 
Further details of our sampling strategy for the fully frustrated ladder can 
be found in Ref.~\cite{PhysRevB.93.054408}. The magnetic susceptibility, 
$\chi(T)$, and specific heat, $C(T)$, are estimated in the usual way for 
the SSE technique, namely from the fluctuations of the total magnetization
and of the expansion order \cite{PhysRevB.68.094423}, respectively.

\begin{table}[t!]
\begin{center}
\begin{tabular}{ | c || c | c | c | c | c | c | c |}
\hline
& $T_i$ & $T^z_{i}$ & $T^+_{i}$       &  $T^-_{i}$        &  $D^z_{i}$  
& $D^+_{i}$        & $D^-_{i}$ \\
\hline
$| S \rangle_{i}$ & 0     & 0  & 0   &  0  &  $| 0 \rangle_{i}$   & $-\sqrt{2}|
 + \rangle_{i}$            & $\sqrt{2}| - \rangle_{i}$ \\
$| 0 \rangle_{i}$ & 1           & 0     & $\sqrt{2}| + \rangle_{i}$          
&  $\sqrt{2}| - \rangle_{i}$    &  $| S \rangle_{i}$   & 0       & 0 \\
$| + \rangle_{i}$ & 1     & 1     & 0           &  $\sqrt{2}| 0 \rangle_{i}$ 
&  0        & 0            & $-\sqrt{2}| S \rangle_{i}$ \\
$| - \rangle_{i}$ & 1           & $-1$    & $\sqrt{2}| 0 \rangle_{i}$    &  0  
&  0          & $\sqrt{2}| S \rangle_{i}$            & 0 \\
\hline
\end{tabular}
\end{center}
\caption{Action of local total-spin and spin-difference operators
(\ref{eq:defTD}) on the local spin-dimer basis states (\ref{eq:sdb}).
Because $\vec{T}^2_{i}$ and $T^z_{i}$ are diagonal in this basis, we quote 
only the corresponding quantum numbers.}
\label{tab:matele}
\end{table}

\subsubsection{Partially frustrated ladder}

\label{sec:SignGen}

For the partially frustrated ladder, the fact that $J_\times \ne J_\|$
mandates working with the more general Hamiltonian of Eq.~(\ref{eq:exehGen}),
where the sign problem returns even in the rung basis. We consider the local 
basis states on rung $i$, 
\begin{eqnarray}
|S \rangle_{i} & = & {\textstyle \frac{1}{\sqrt{2}}} ( |\uparrow \downarrow 
\rangle_{i} - |\downarrow \uparrow \rangle_{i}) \, , \nonumber \\
|0 \rangle_{i} & = & {\textstyle \frac{1}{\sqrt{2}}} ( |\uparrow \downarrow 
\rangle_{i} + |\downarrow \uparrow \rangle_{i}) \, , \qquad
|+ \rangle_{i} = |\uparrow \uparrow \rangle_{i}, \qquad
|- \rangle_{i} = |\downarrow \downarrow \rangle_{i} \, ,
\label{eq:sdb}
\end{eqnarray}
in terms of which the matrix elements of the operators (\ref{eq:defTD})
are given in Table \ref{tab:matele}. In the SSE framework, the specific 
requirement is that the matrix elements of $-H$ should be non-negative
\cite{PhysRevE.66.046701}. This can always be ensured for diagonal matrix
elements by addition of a suitable global constant to the Hamiltonian,
which only shifts the zero of energy and otherwise has no effect on the 
physics. If the rewritten Hamiltonian (\ref{eq:exehGen}) is defined on a 
bipartite lattice, as is the case for the ladder, one may further perform 
a $\pi$-rotation around the $z$-axis on one of the two sublattices, 
\begin{equation}
T_{2\,i}^\pm \to -T_{2\,i}^\pm \, , \qquad
D_{2\,i}^\pm \to -D_{2\,i}^\pm  \, ,
\label{eq:piRot}
\end{equation}
which preserves the commutation relations (\ref{eq:TDcomm}) but renders the 
sign of the $T_i^\pm\,T_{i+1}^\mp$ terms negative, as required.\footnote{This 
sublattice rotation ensures that all of the corresponding configurations 
always have positive weight and thus need not actually be carried out 
explicitly; one source for further details is Ref.~\cite{EvertzLoop03}.}

The prefactor specifying the sign of the $\vec{D} \cdot \vec{D}$ interaction 
in Eq.~(\ref{eq:exehGen}) is positive for the case $J_\times < J_\|$ considered 
here. Although the $D^z\,D^z$ terms are off-diagonal and non-negative in 
the basis (\ref{eq:sdb}), as Table \ref{tab:matele} makes clear, they pose 
no problem here \cite{PhysRevB.93.054408,PhysRevLett.117.197203}. One way 
to see this explicitly is to send $\vec{D}_{2i} \to -\vec{D}_{2i}$ for one 
sublattice, which corresponds to an interchange $1 \leftrightarrow 2$ of 
the two legs on every second rung in the transformation (\ref{eq:defTD}). 
This changes the sign of the $D^z_i \, D^z_{i+1}$ terms such that they can 
also be considered to be negative.

These considerations leave only the $D_i^\pm \, D_{i+1}^\mp$ terms. Inspection 
of Table \ref{tab:matele} shows that their matrix elements can be both 
positive and negative, such that they do actually give rise to a sign 
problem. However, the $D_i^\pm \, D_{i+1}^\mp$ terms also exchange a pair of 
local quantum numbers, ($T_i$, $T_{i+1}$). Because such an exchange must be 
compensated by further similar terms in the SSE operator string, the 
occurrence of this type of term is severely restricted. The resulting 
sign problem therefore turns out to be remarkably mild, as we will 
demonstrate in our numerical results.

We comment finally that the absence of local conservation laws improves 
the ergodicity of the Monte Carlo sampling. Thus we found that it is not 
necessary to employ parallel tempering for simulations with any parameter 
sets where $J_\times \ne J_\|$.

\begin{figure}[t!]
\centering\includegraphics[width=0.75\columnwidth]{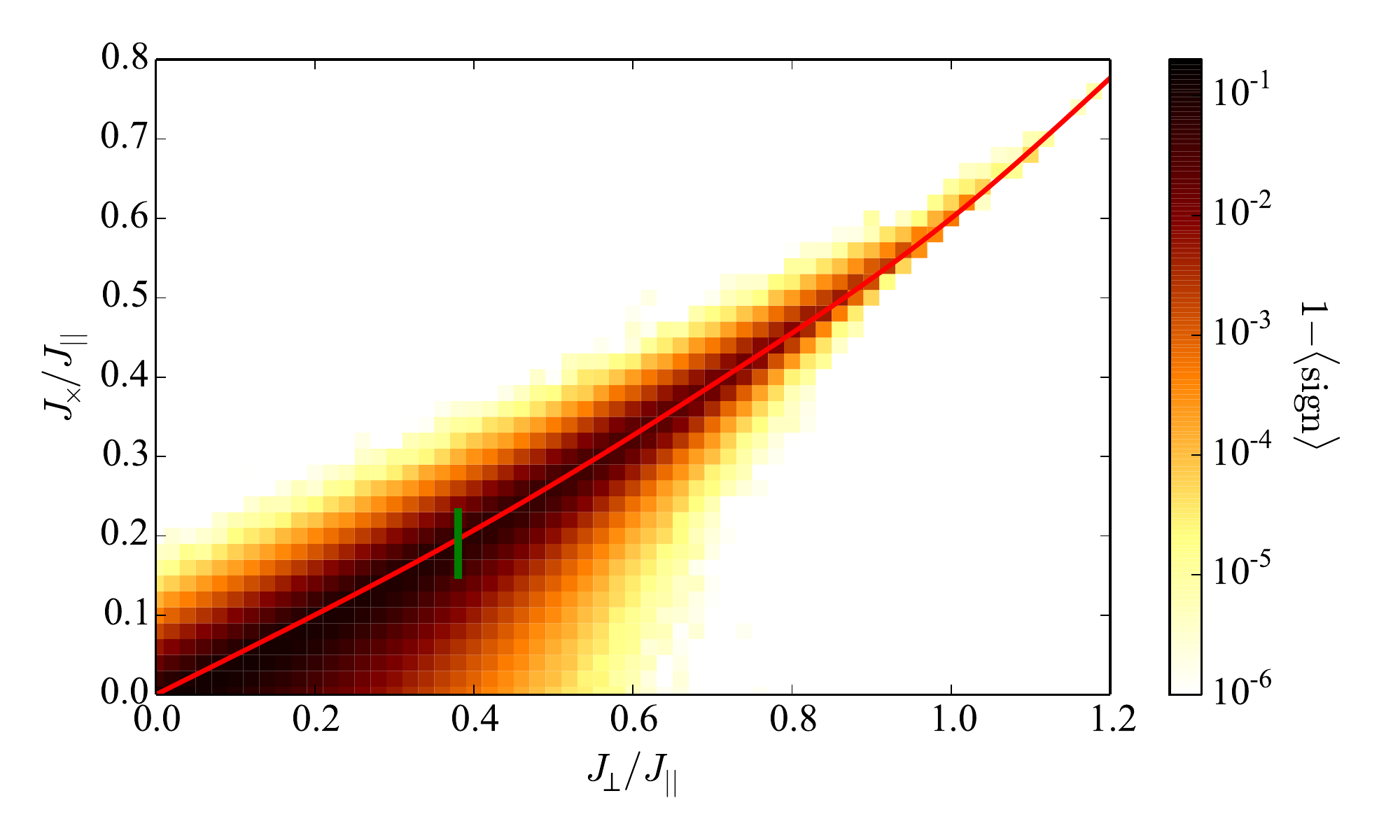}
\caption{Deviation of the average sign from one, $1 - \langle \text{sign} 
\rangle$, computed by QMC simulations in the rung basis using ladders of 
$L = 50$ rungs at $T/J_\| = 0.05$. The solid red line is the phase transition 
from Fig.~\ref{fig:ladder_phaseDiag} and the vertical green line shows the 
scan direction detailed in the inset of Fig.~\ref{fig:signJR0_38JD0_196}.
White regions correspond to a deviation in average sign of less than $10^{-6}$;
outside the range of parameters covered by the figure, $\langle \text{sign} 
\rangle$ is numerically indistinguishable from 1.}
\label{fig:signScan}
\end{figure}

\subsection{Numerical results for the average sign}

\label{sec:SignNum}

Because the sign problem is usually worst at low temperatures
\cite{PhysRevLett.94.170201}, we have performed a scan to compute the 
average sign, $\langle \text{sign}\rangle$, throughout the parameter space
for a fixed low temperature, $T/J_\| = 0.05$, for ladders of $L = 50$ rungs.
We note that $T/J_\| = 0.05$ is so low that, for the case shown in 
Fig.~\ref{fig:signJR0_38JD0_196site}, it would be impossible to obtain 
any meaningful results by performing simulations in the single-site basis. 
By contrast, the results for simulations performed in the rung basis,
shown in Fig.~\ref{fig:signScan}, are remarkably well-behaved.

In detail, the average sign is indistinguishable from unity over the majority 
of the phase diagram, and a noticeable deviation appears only within the 
transition region from the rung-triplet to the rung-singlet phase. Although 
this deviation is largest in the low-$J_\perp$ region ($J_\perp, J_\times \ll 
J_\|$), we stress that $\langle \text{sign} \rangle > 0.86$ over the entire 
parameter regime for simulations with $L = 50$ at $T/J_\| = 0.05$. 
Qualitatively, the maximum of the difference $1 - \langle \text{sign} 
\rangle$ in Fig.~\ref{fig:signScan} traces the phase transition line
(Fig.~\ref{fig:ladder_phaseDiag}).

\begin{figure}[t!]
\centering\includegraphics[width=0.66\columnwidth]{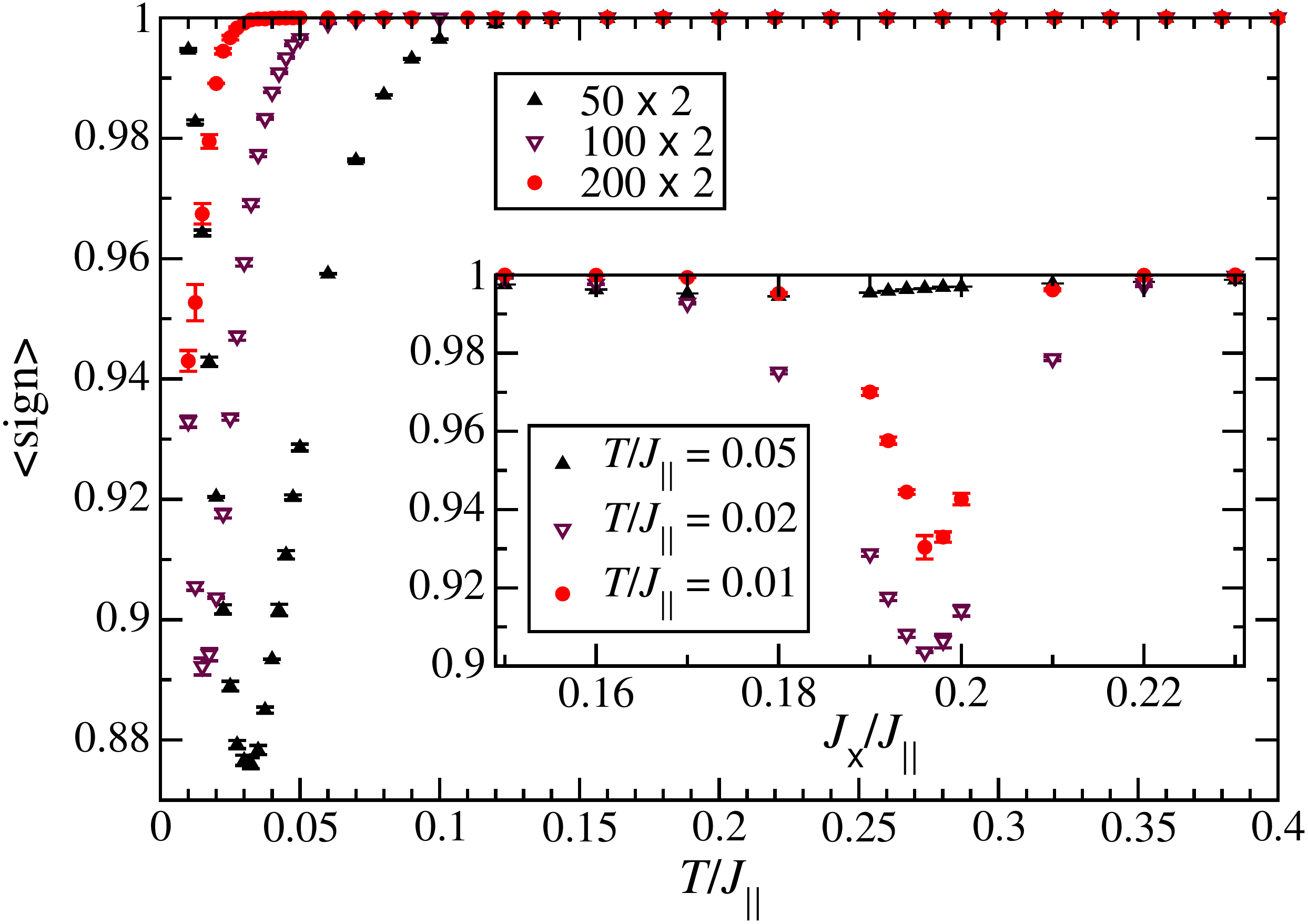}
\caption{Average sign, $\langle \text{sign} \rangle$, for a QMC simulation
in the rung basis using the parameters $J_\perp = 0.38$, $J_\| = 1$, and 
$J_\times = 0.196$, shown as a function of temperature for ladders with 
$L = 50$, $100$, and $200$ rungs. The thermodynamic response of this 
system is shown in Fig.~\ref{fig:thermoBreakJR0_38JD0_196}. Inset: 
$\langle \text{sign} \rangle$ shown as a function of $J_\times$ for a 
ladder with $J_\| = 1$, $J_\perp = 0.38$, and $L = 100$ rungs, at three 
temperatures, $T/J_\| = 0.01$, $0.02$, and $0.05$.}
\label{fig:signJR0_38JD0_196}
\end{figure}

The behavior of the average sign is investigated more closely for a ladder 
with $J_\perp/J_\| = 0.38$ in Fig.~\ref{fig:signJR0_38JD0_196}. The inset  
shows QMC data for $\langle \text{sign} \rangle$ as a function of $J_\times
 / J_\|$ at various low temperatures. As noted in Sec.~\ref{sec:Model}, 
this transition region between the rung-triplet and \hbox{-singlet} phases 
has been studied intensively, in particular with regard to a possible 
intermediate dimerized phase \cite{PhysRevB.73.224433,PhysRevB.77.205121, 
PhysRevB.77.214418,PhysRevB.86.075133,Chen2016}. While a rather sharp dip 
can be identified in $\langle \text{sign} \rangle$ for $J_\times/J_\| \approx 
0.2$ in the data for $T/J_\| = 0.02$ and 0.01, the absolute value of $\langle 
\text{sign} \rangle$ remains above $0.9$ at all times, meaning that QMC 
simulations performed in this regime are fully reliable. Surprisingly, the 
average sign shows a non-monotonic temperature-dependence, increasing at 
the lowest temperature considered here, $T/J_\| = 0.01$, compared to $T/J_\|
 = 0.02$. The position of the minimum in $\langle \text{sign} \rangle$ at 
sufficiently low temperatures appears to constitute a signature of the phase 
transition that is comparable in accuracy with dedicated $T = 0$ methods
\cite{PhysRevB.73.224433,PhysRevB.77.205121,PhysRevB.77.214418,
PhysRevB.86.075133,Chen2016}, and thus we obtain the estimate $J_{\times,c}
 / J_\| \approx 0.196$ at  $J_\perp/J_\| = 0.38$.

The main panel of Fig.~\ref{fig:signJR0_38JD0_196} shows $\langle \text{sign} 
\rangle$ as a function of temperature for three different system sizes at the 
estimated critical value, $J_{\times,c}/J_\| \approx 0.196$ for $J_\perp/J_\| = 0.38
$. In contrast to the single-site basis (Fig.~\ref{fig:signJR0_38JD0_196site}), 
we observe deviations of the average sign from unity only for temperatures
$T/J_\| < 0.1$. Our results display not only the aforementioned non-monotonic 
temperature-dependence, whereby $\langle \text{sign} \rangle$ increases 
towards lower temperatures, but also that, quite unlike 
Fig.~\ref{fig:signJR0_38JD0_196site}, the average sign increases with 
increasing system size. 

The fact that the average sign is close to unity may be expected deep
inside the rung-singlet and \hbox{-triplet} phases, where the respective ground 
states are thought to be well captured by the rung basis. Only on approaching 
the phase transition, and indeed only at rather weak $J_\perp/J_\|$, do we 
observe appreciable deviations from unity, indicating that inter-rung 
fluctuations are large. Still, with values $\langle \text{sign} \rangle 
\gtrsim 0.9$, the average sign remains remarkably well-behaved throughout 
the phase diagram. In particular, the fact that its behavior improves with 
decreasing temperature and with increasing system size is in sharp contrast 
to expectations \cite{PhysRevLett.94.170201} and to the situation in the 
single-site basis (Fig.~\ref{fig:signJR0_38JD0_196site}), and offers still 
greater potential for minimizing the sign problem. 

To account for this behavior at a qualitative level, in 
Subsec.~\ref{sec:SignGen} we pointed out that configurations with a negative 
sign are severely constrained for the frustrated ladder. Indeed, our empirical 
observation is that the sign problem in the rung basis is completely absent 
for systems with open boundary conditions, while for periodic boundary 
conditions the first term giving rise to a sign problem occurs at order 
$L+1$ and contains an operator string that wraps around the entire system, 
as also noted in Ref.~\cite{PhysRevLett.117.197203}. This is consistent with 
the fact that a significant number of configurations with negative sign arises 
only in a small temperature window, as well as with the fact that such 
configurations never constitute a macroscopic fraction of all configurations.

Despite the complete absence of a sign problem when working in the rung basis 
on a ladder with open boundaries, periodic boundary conditions offer several 
advantages in the calculation of bulk thermodynamic properties for finite 
systems. Not least of these is the elimination of surface terms, which could 
result in some particular problems in the Haldane phase of the ladder. In any 
case, when $\langle \text{sign} \rangle \gtrsim 0.9$ it is no longer true that 
the sign ``problem'' is a limiting factor for QMC simulations, because other 
algorithmic aspects, such as autocorrelation times, become more relevant 
for simulation efficiency. Specifically, if $\langle \text{sign} \rangle 
\approx 0.9$ then it is necessary to collect only 25\%\ more samples in 
order to compensate for the loss of accuracy associated with sign effects. 
We remark in closing that the single-site basis remains nevertheless the 
natural choice for QMC simulations of unfrustrated ladders \cite{Hida91,PhysRevB.47.3196,
PhysRevLett.77.1865,PhysRevB.54.R3714,ladderQMCrev,PhysRevB.93.054408}, and
accordingly we do not pursue the case $J_\times = 0$ here.

\section{Results: Thermodynamic properties}

\label{sec:Results}

\begin{figure}[t!]
\includegraphics[width=0.49\columnwidth]{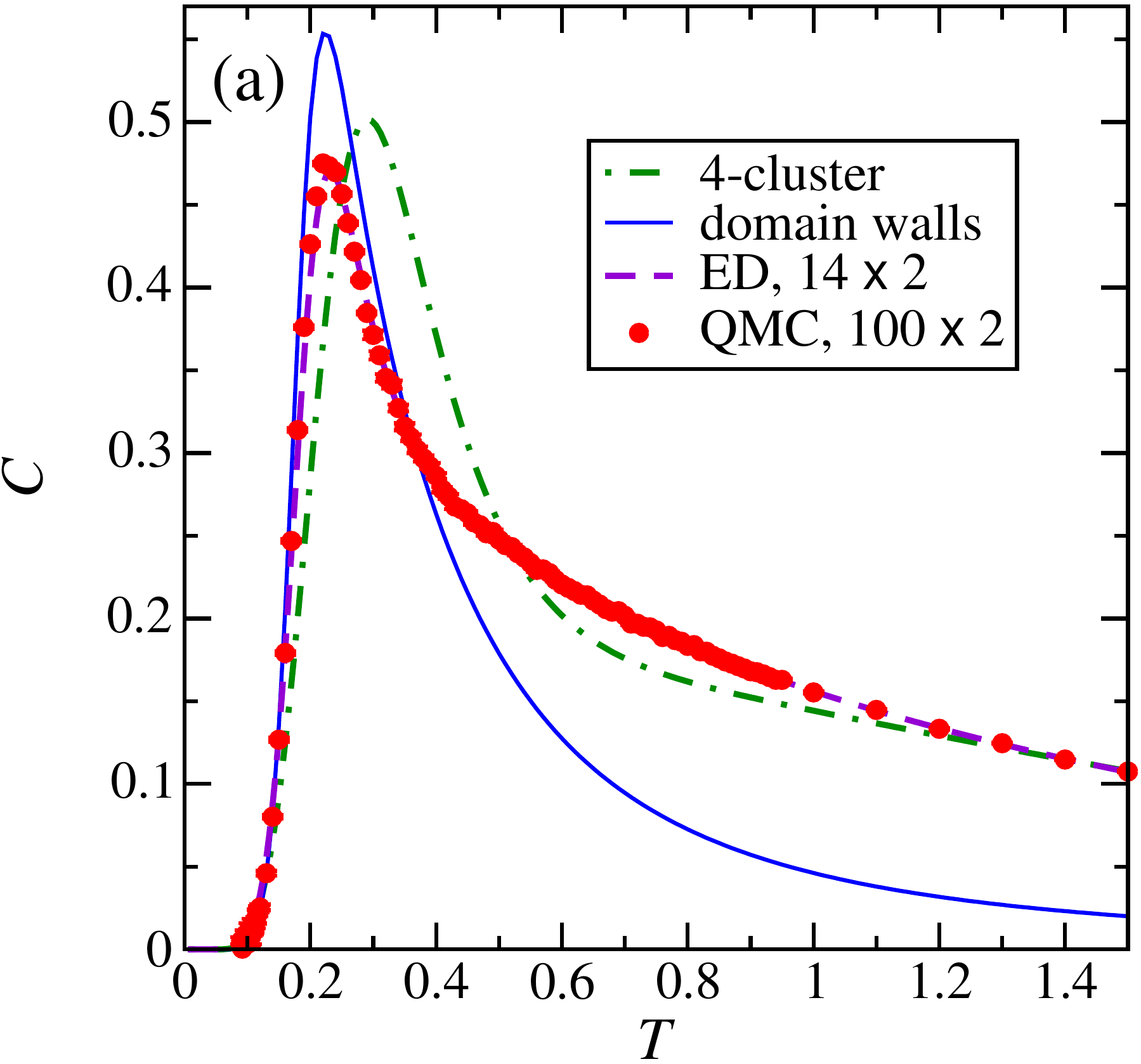}\hfill%
\includegraphics[width=0.49\columnwidth]{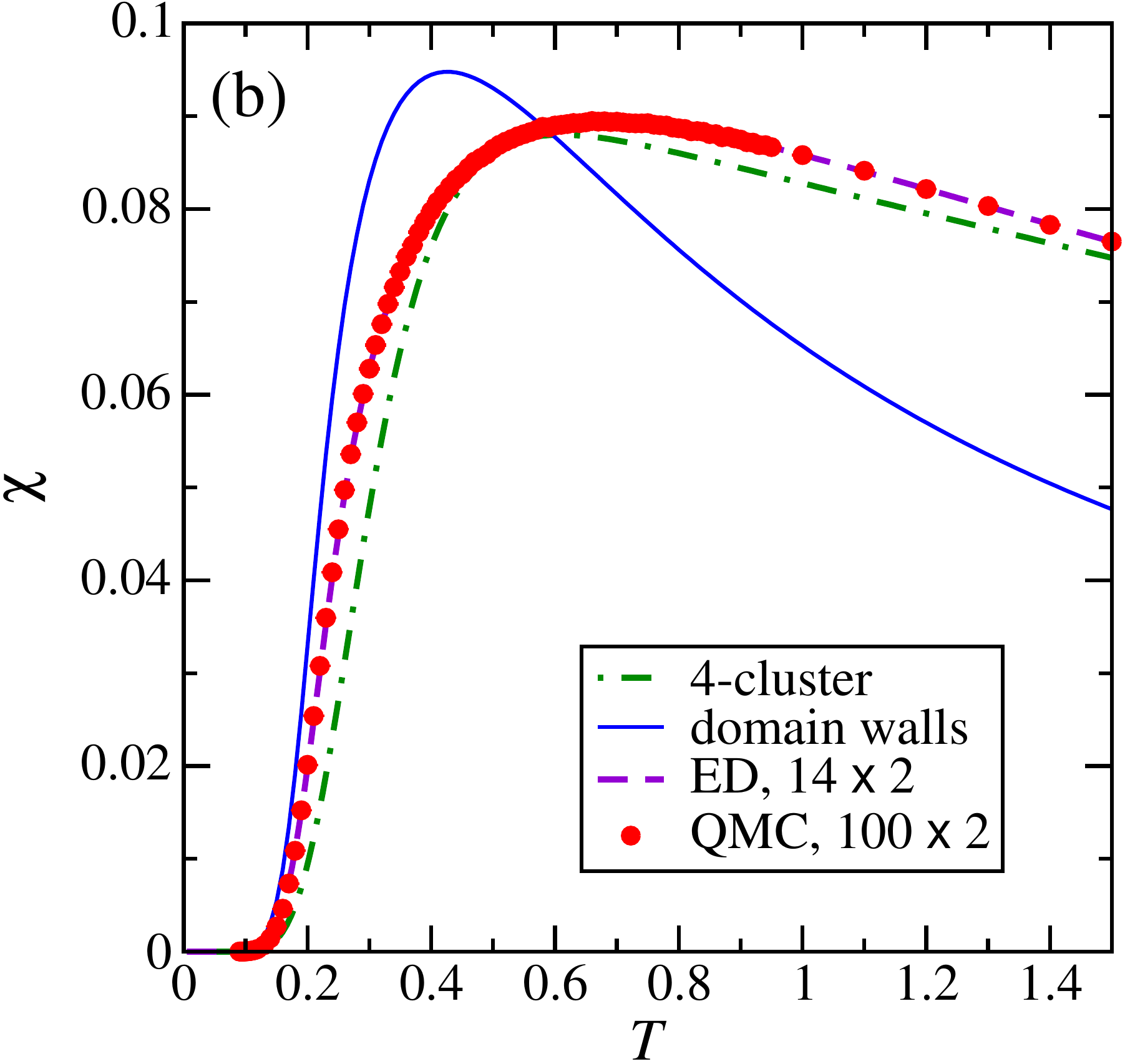}
\caption{Magnetic specific heat, $C$ (left panel), and susceptibility, $\chi$ 
(right panel), shown per spin for a fully frustrated ladder with rung coupling 
$J_\perp = 1.45$ and inter-rung couplings $J_\times = J_\| = 1$. We compare 
QMC results obtained for a ladder of $L = 100$ rungs with ED results for 
$L = 14$ and with approximate calculations based on clusters of 4 rungs 
and on a model of non-interacting domain walls.}
\label{fig:thermoFFL1_45}
\end{figure}

In Figs.~\ref{fig:thermoFFL1_45}--\ref{fig:thermoBreakJR0_38JD0_196} we 
present numerical results for the magnetic specific heat and susceptibility 
of frustrated spin ladders for a selection of representative points in the 
phase diagram, as shown by the corresponding numbered circles in 
Fig.~\ref{fig:ladder_phaseDiag}. In these figures, the sizes of the QMC 
error bars are in general significantly smaller than the symbol sizes, 
demonstrating the high quality of the nearly-sign-free simulations. For 
interpretive purposes we include certain analytical results where appropriate 
and for numerical comparison we also include exact diagonalization (ED) 
results obtained for small systems.\footnote{ED results are based on the 
Hamiltonian of Eq.~(\ref{eq:essh}) whenever $J_\times \neq J_\|$. With the 
help of the spatial symmetries and conservation of total $S^z$, it is 
straightforward to perform full diagonalization of a system with up to 20 
spins 1/2 ($L \le 10$ rungs). At $J_\times = J_\|$, represented here by 
Figs.~\ref{fig:thermoFFL1_45} and \ref{fig:thermoFFL1_4}, one may take 
advantage of the additional local conservation laws to further simplify 
the problem \cite{LTP07,azurite11,PhysRevB.86.054412,PhysRevB.93.054408}. 
Although the complexity remains exponential in $L$, the rewriting 
(\ref{eq:exeh}) makes it possible to access $L = 14$ rungs (28 $S = 1/2$ 
spins) \cite{PhysRevB.93.054408}.} At points sufficiently deep inside the 
rung-singlet phase, finite-size effects are negligible at all temperatures. 
In these cases, and at all high temperatures, the ED results coincide exactly 
with the QMC ones, validating again the reliability of the simulations.

\subsection{Fully frustrated ladder}

\label{sec:FFL}

We begin by presenting for reference two examples of the thermodynamic 
response of fully frustrated ladders ($J_\times = J_\|$), a case already 
investigated in some detail in Ref.~\cite{PhysRevB.93.054408}. We choose 
first a point in the rung-singlet phase but close to the phase transition,
$J_\perp/J_\| = 1.45$, and present numerical results for the specific heat 
[Fig.~\ref{fig:thermoFFL1_45}(a)] and susceptibility 
[Fig.~\ref{fig:thermoFFL1_45}(b)]. On a technical level, it is important
to note that the ED results for $L = 14$ and QMC for $L = 100$ rungs are 
in good agreement for both quantities, with discrepancies appearing only 
in the specific heat and only around its maximum. Thus we conclude that 
the QMC results for $L = 100$ rungs in Fig.~\ref{fig:thermoFFL1_45} are
indistinguishable from the thermodynamic limit (indeed, the ED data for 
$L = 14$ already yield a good approximation to the infinite system). 

\begin{figure}[t!]
\includegraphics[width=0.49\columnwidth]{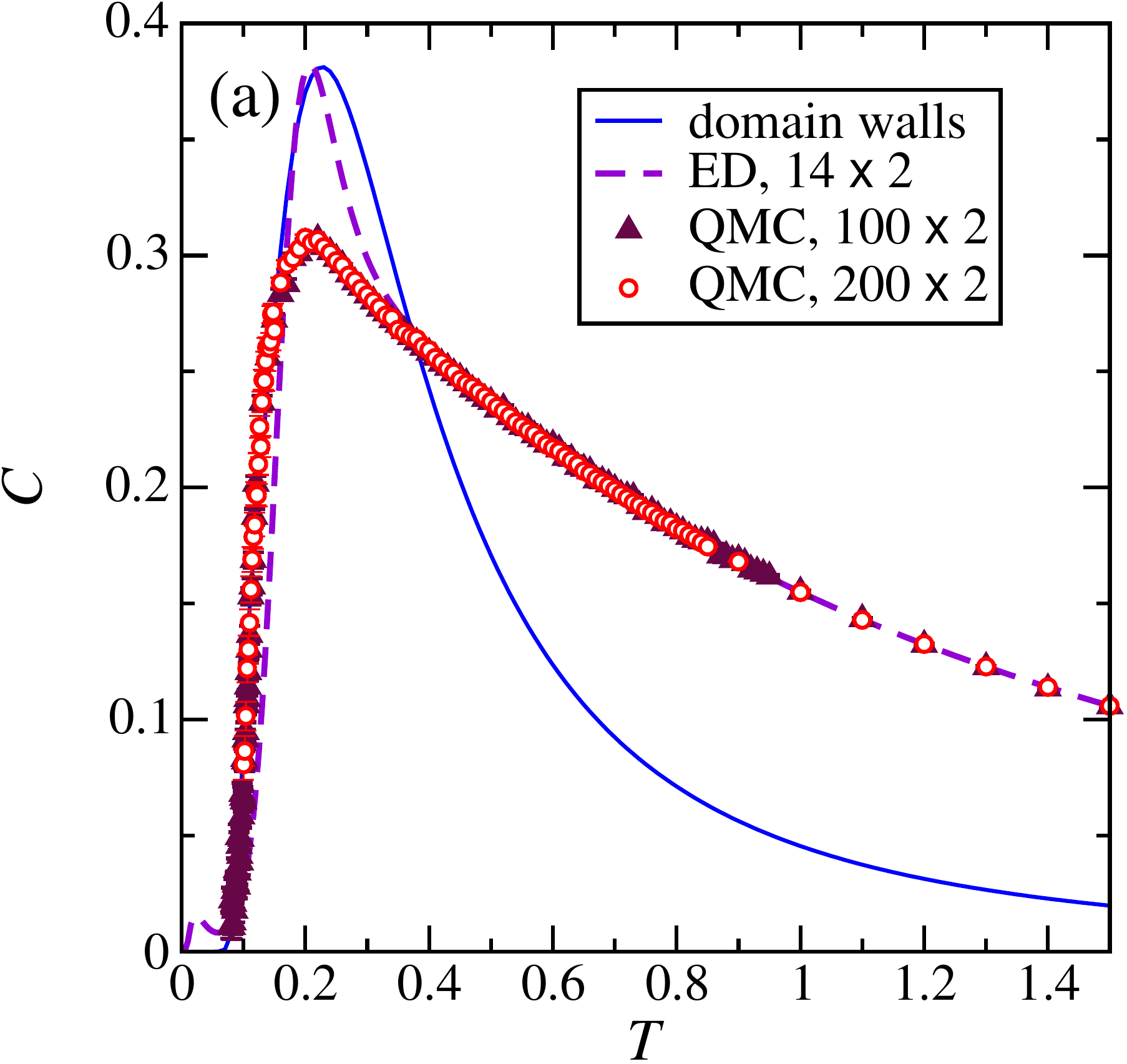}\hfill%
\includegraphics[width=0.49\columnwidth]{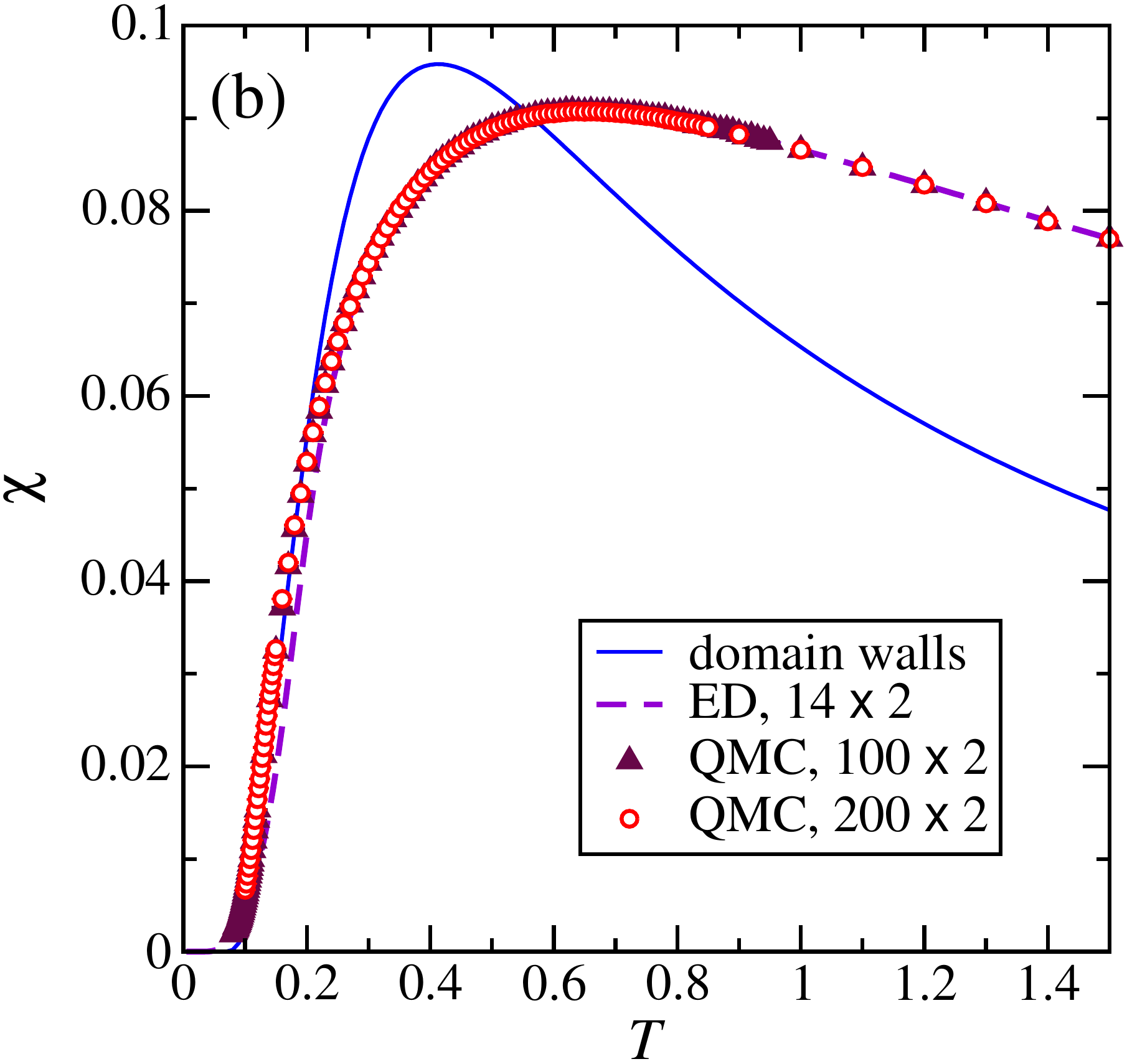}
\caption{Magnetic specific heat, $C$, and susceptibility, $\chi$, shown per 
spin for a fully frustrated ladder with rung coupling $J_\perp = 1.4$ and 
inter-rung couplings $J_\times = J_\| = 1$. These values lie very close to 
the critical line in Fig.~\ref{fig:ladder_phaseDiag}. We compare QMC results 
obtained for ladders of $L = 100$ and 200 rungs with ED results for $L = 14$ 
and with a model of non-interacting domain walls.}
\label{fig:thermoFFL1_4}
\end{figure}

$\chi(T)$ in Fig.~\ref{fig:thermoFFL1_45}(b) exhibits a broad maximum, 
which is characteristic for highly frustrated magnets. Its temperature, 
$T_{\rm max}^\chi \simeq 0.68\,J_\|$, is not particularly low in comparison 
to the one-triplon energy, $\tilde{E}_{n=1} = J_\perp = 1.45\,J_\|$. However, 
as discussed in Ref.~\cite{PhysRevB.93.054408}, a characteristic quantity 
more useful than the broad maximum is that $\chi$ attains half of its maximal 
value at the comparatively low temperature $T_{\rm half}^\chi \simeq 0.248\,J_\|$. 
The rise of the specific heat in Fig.~\ref{fig:thermoFFL1_45}(a) occurs at an 
even lower temperature, $T_{\rm half}^C \simeq 0.169\,J_\|$, in accord with the 
fact that $C(T)$ is sensitive to the lowest singlet bound state, which appears 
at $\tilde{E}_{n=2}^1 = 0.9\,J_\|$ for these parameters. However, the most 
striking feature of Fig.~\ref{fig:thermoFFL1_45}(a) is clearly the emergence 
of a remarkably sharp maximum in $C(T)$ at the very low temperature 
$T_{\rm max}^C \simeq 0.231\,J_\|$.

These results were interpreted in Ref.~\cite{PhysRevB.93.054408} by a 
detailed analysis of the many rung-triplet bound states in the rung-singlet 
phase, which move to anomalously low energies near the quantum phase 
transition. For illustration, we appeal to two analytical approximations 
that yield the other two curves included in each panel of 
Fig.~\ref{fig:thermoFFL1_45}. First, a computation taking into account all 
excited states of $n$-triplon clusters, with $n \le 4$, yields an accurate 
description deep in the rung-singlet phase \cite{PhysRevB.93.054408} and 
also a good account of the high-temperature behavior at a point as close 
to the transition as $J_\perp/J_\| = 1.45$. However, it cannot provide an 
accurate reproduction of either the sharp nature or the low effective 
temperature scale of the specific-heat peak. For this, a ``domain-wall'' 
model \cite{PhysRevB.93.054408} yields a better description, particularly 
of the low-temperature onset ($T_{\rm half}^C$) and maximum position ($T_{\rm 
max}^C$) in $C(T)$. Near the transition, these walls exist between domains 
of almost-degenerate rung-singlet and \hbox{-triplet} states, and describe 
the contributions of $S = 1/2$ end-spins terminating the $n$-site rung-triplet 
(spin-1) chain segments in a rung-singlet background for all values of $n$ 
\cite{PhysRevLett.59.799,Kennedy90,NUZ97}. We conclude that the characteristic 
emergent temperature scales ($T_{\rm half}^C$, $T_{\rm max}^C$, and $T_{\rm half}^\chi$)
reflect not only the energies of the low-lying excited states but also their 
degeneracies, and can therefore be considered as signatures of the complex 
spectrum of bound states. 

\begin{figure}[t!]
\includegraphics[width=0.49\columnwidth]{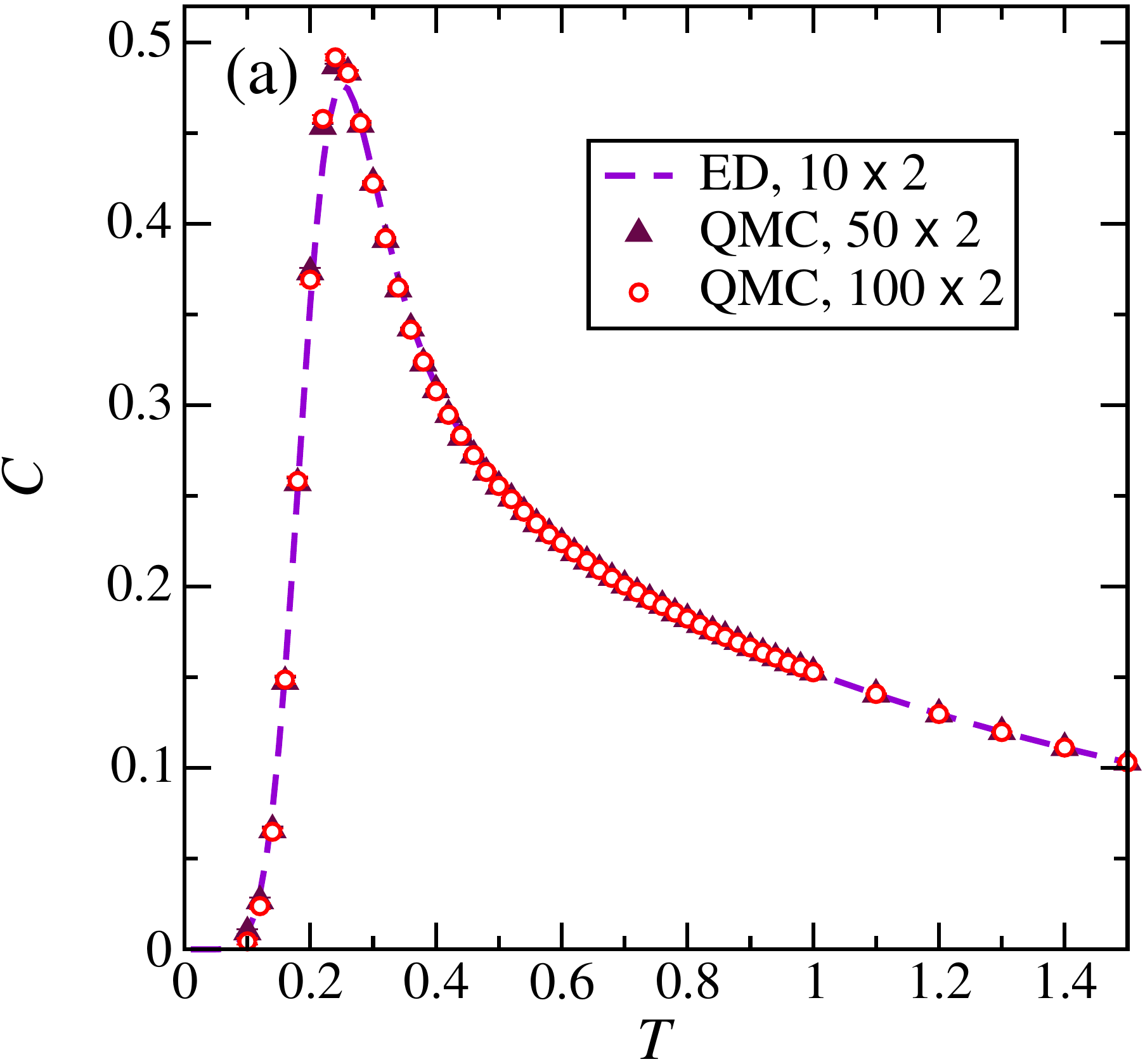}\hfill%
\includegraphics[width=0.49\columnwidth]{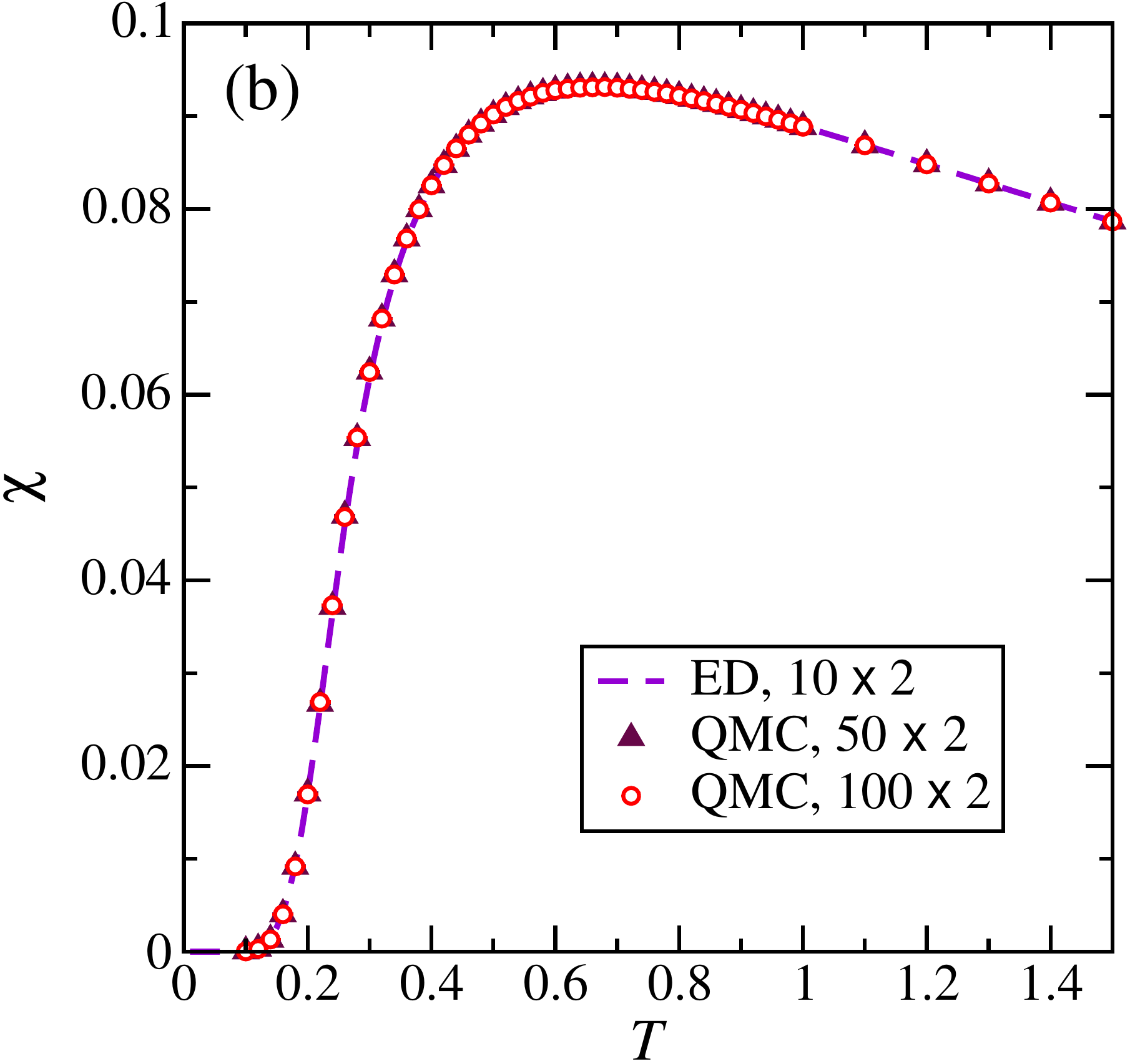}
\caption{Magnetic specific heat, $C$, and susceptibility, $\chi$, shown per 
spin for a partially frustrated ladder with rung coupling $J_\perp = 1.4$, 
leg coupling $J_\| = 1$, and diagonal coupling $J_\times = 0.9$. We compare 
QMC results obtained for ladders of $L = 50$ and 100 rungs with ED results 
for $L = 10$.} 
\label{fig:thermoBreakJR1_4JD0_9}
\end{figure}

Figure \ref{fig:thermoFFL1_4} presents thermodynamic results for the point
$J_\perp/J_\| = 1.4$, which lies effectively at the phase transition for 
$J_\times = J_\|$ (Fig.~\ref{fig:ladder_phaseDiag}). We observe that the QMC 
data for $L = 100$ and $200$ rungs are very close, showing that $L = 100$ 
can be considered as fully representative of the thermodynamic limit, even 
at the phase transition itself. Nonetheless, finite-size effects visible in 
the ED results are significantly stronger than in Fig.~\ref{fig:thermoFFL1_45}. 
In particular, the $L = 14$ curve in Fig.~\ref{fig:thermoFFL1_4}(a) suggests a 
sharp low-temperature maximum in $C(T)$, whereas the QMC results demonstrate 
that in fact only a shoulder survives in the thermodynamic limit at the 
critical coupling (as a consequence of contributions from $n$-rung triplet 
clusters of all length scales \cite{PhysRevB.93.054408}).

The characteristic temperatures observed at $J_\perp/J_\| = 1.45$ 
are further reduced at $J_\perp/J_\| = 1.4$, with (from the QMC data in 
Fig.~\ref{fig:thermoFFL1_4}) $T_{\rm max}^C \simeq 0.199\,J_\|$, $T_{\rm half}^C 
\simeq 0.112\,J_\|$, and $T_{\rm half}^\chi \simeq 0.179\,J_\|$, reflecting a 
further reduction of the relevant excitation energies \cite{PhysRevB.93.054408}.
While the short-cluster approximation is not well suited to this case, the 
domain-wall model continues to yield an accurate description at least of 
the low-temperature onset of both $C(T)$ and $\chi(T)$, demonstrating the 
importance of these walls as the effective lowest-energy excitations.

\subsection{Highly frustrated ladder}

\label{sec:High}

We extend our considerations by exploring the broader regions of the phase 
diagram (Fig.~\ref{fig:ladder_phaseDiag}). To assess the influence of the 
local conservation laws effective at $J_\times = J_\|$, we first move only a 
little away from full frustration. We consider the illustrative case $J_\perp/
J_\| = 1.4$, $J_\times/J_\| = 0.9$, which lies in the rung-singlet phase but 
remains close to the transition. Figure~\ref{fig:thermoBreakJR1_4JD0_9} 
presents numerical results for $C(T)$ and $\chi(T)$ of this still highly 
frustrated ladder. The QMC procedure is now based on the expression of 
Eq.~(\ref{eq:exehGen}), and while a sign problem is in general present when 
$J_\times \ne J_\|$, the average sign remains numerically indistinguishable 
from $1$ (Sec.~\ref{sec:SignNum}). In Fig.~\ref{fig:thermoBreakJR1_4JD0_9}
we include ED results for $L = 10$ rungs, which exhibit only minor finite-size 
effects around the maximum of $C(T)$ at $T/J_\| \approx 0.25$. The QMC data 
for $L = 50$ can definitely be regarded as representative of the thermodynamic 
limit.

\begin{figure}[t!]
\includegraphics[width=0.49\columnwidth]{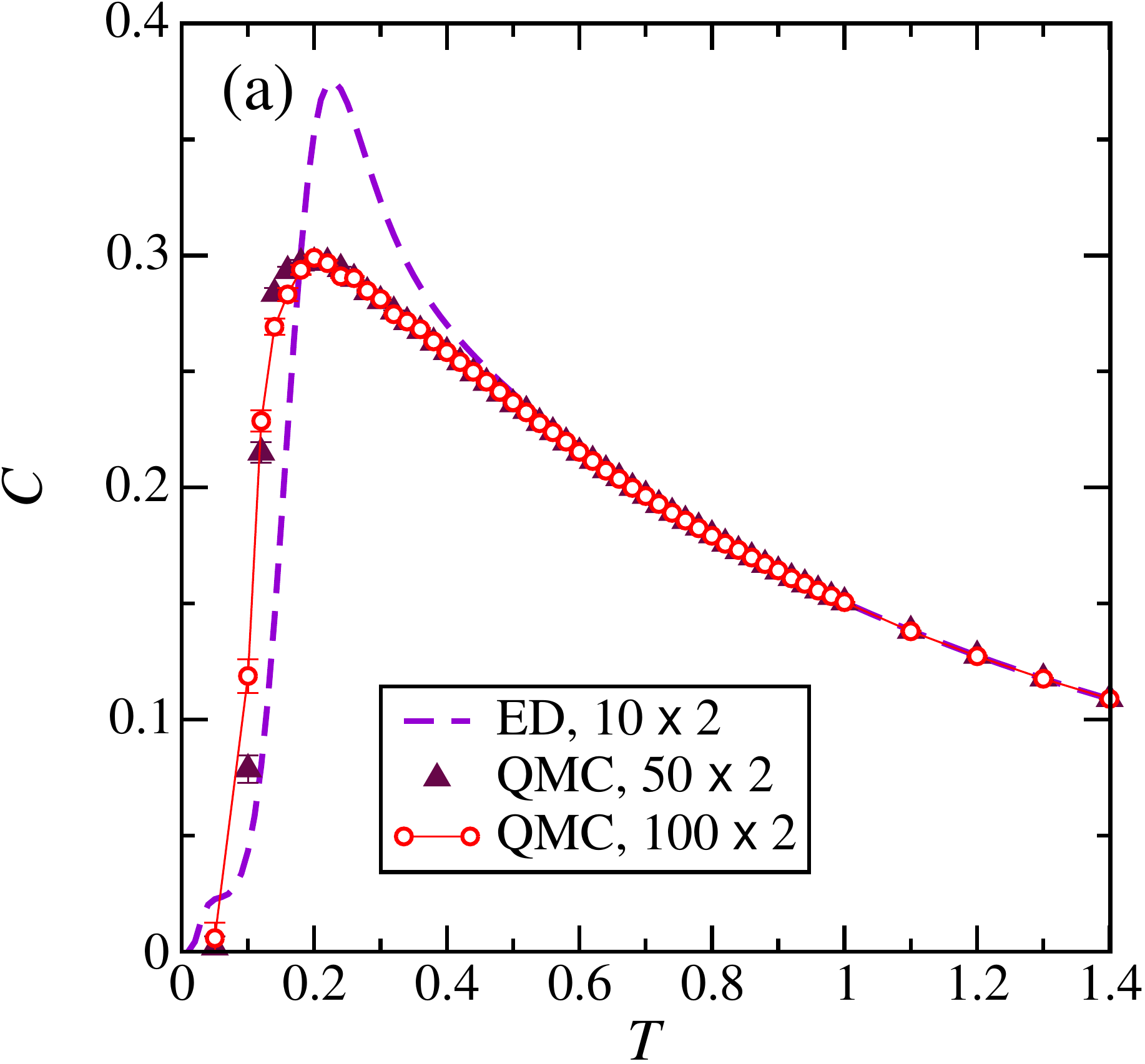}\hfill%
\includegraphics[width=0.49\columnwidth]{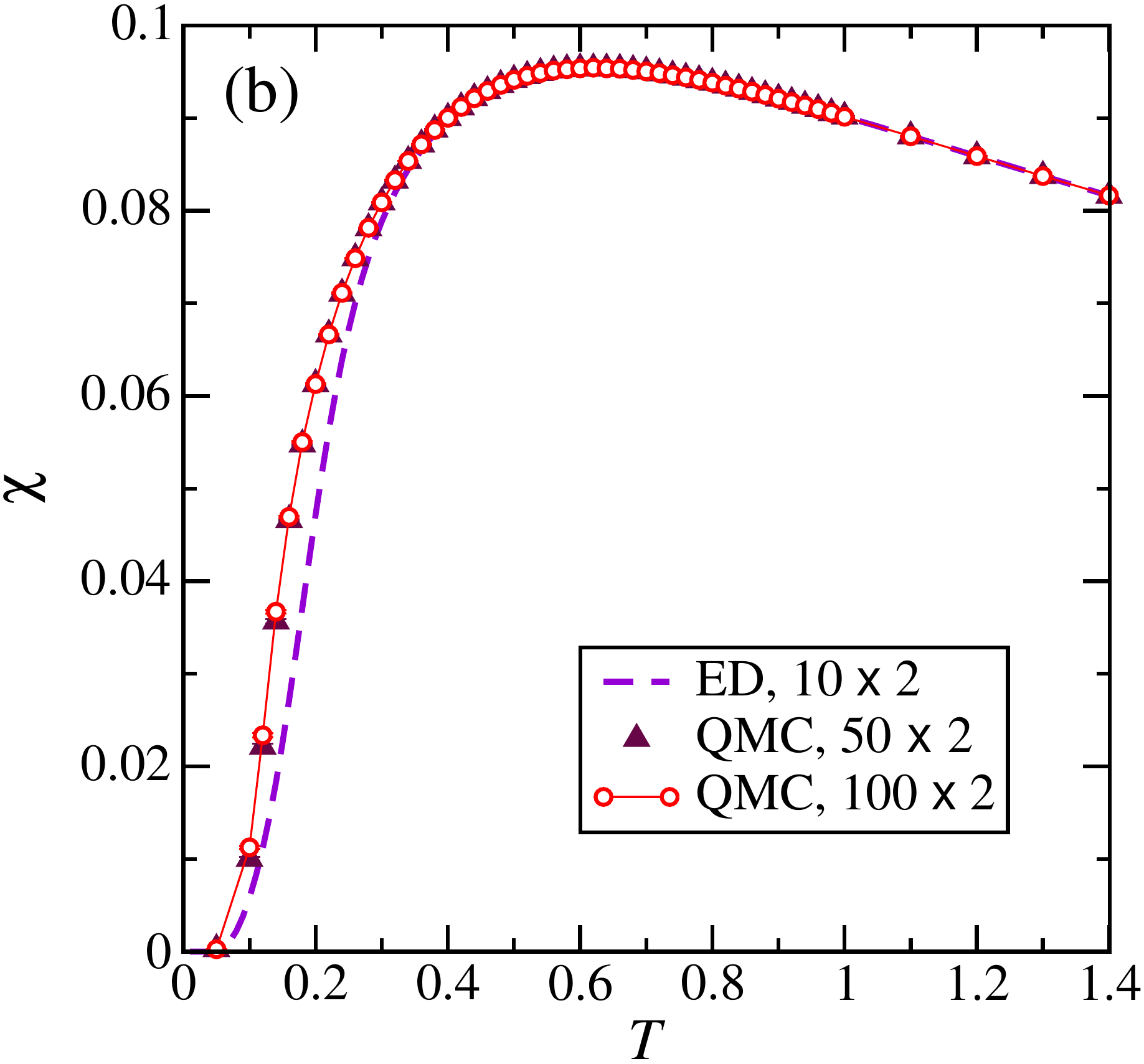}
\caption{Magnetic specific heat, $C$, and susceptibility, $\chi$, shown 
per spin for a ladder with rung coupling $J_\perp = 1.32$, leg coupling 
$J_\| = 1$, and diagonal coupling $J_\times = 0.9$. We compare QMC results 
obtained for ladders of $L = 50$ and 100 rungs with ED results for $L = 10$. 
The line connecting the $L = 100$ QMC data points is a guide to the eye.}
\label{fig:thermoBreakJR1_32JD0_9}
\end{figure}

The parameters of Fig.~\ref{fig:thermoBreakJR1_4JD0_9} were chosen to give 
a point approximately the same distance from the phase-transition line as 
that shown for a fully frustrated ladder in Fig.~\ref{fig:thermoFFL1_45}.
The overall qualitative features are very similar, including in particular
the low onset temperatures and the low-temperature maximum in $C(T)$. This 
result demonstrates that the qualitative features of the spectrum of 
multi-triplet bound states \cite{PhysRevB.93.054408}, and its evolution 
on approaching the transition, are not strongly affected by the imperfect 
frustration. Thus while the fully frustrated line and the presence of local 
conservation laws are helpful for an analytical understanding, including of 
the behavior at all finite temperatures, they do not lead to any unique 
physical properties not present in the more general frustrated ladder.

Figure \ref{fig:thermoBreakJR1_32JD0_9} presents the example $J_\perp/J_\|
 = 1.32$, $J_\times/J_\| = 0.9$, which is again close to full frustration and 
lies essentially on the first-order transition between the rung-singlet and 
\hbox{-triplet} phases (Fig.~\ref{fig:ladder_phaseDiag}). The thermodynamic 
response, including the low onset temperatures, a shoulder-type feature in 
the specific heat at low temperatures, and even the extent of finite-size 
effects, is very similar to the fully frustrated case shown in 
Fig.~\ref{fig:thermoFFL1_4}. This reinforces the observation that 
the local conservation laws present at $J_\times = J_\|$ are useful for 
interpretation but are not essential in determining the qualitative 
thermodynamic features of the frustrated ladder, at least in the regime 
of small ``detuning'' $J_\times \ne J_\|$.

\subsection{Rung-singlet phase}

\label{sec:RungSinglet}

\begin{figure}[t!]
\includegraphics[width=0.49\columnwidth]{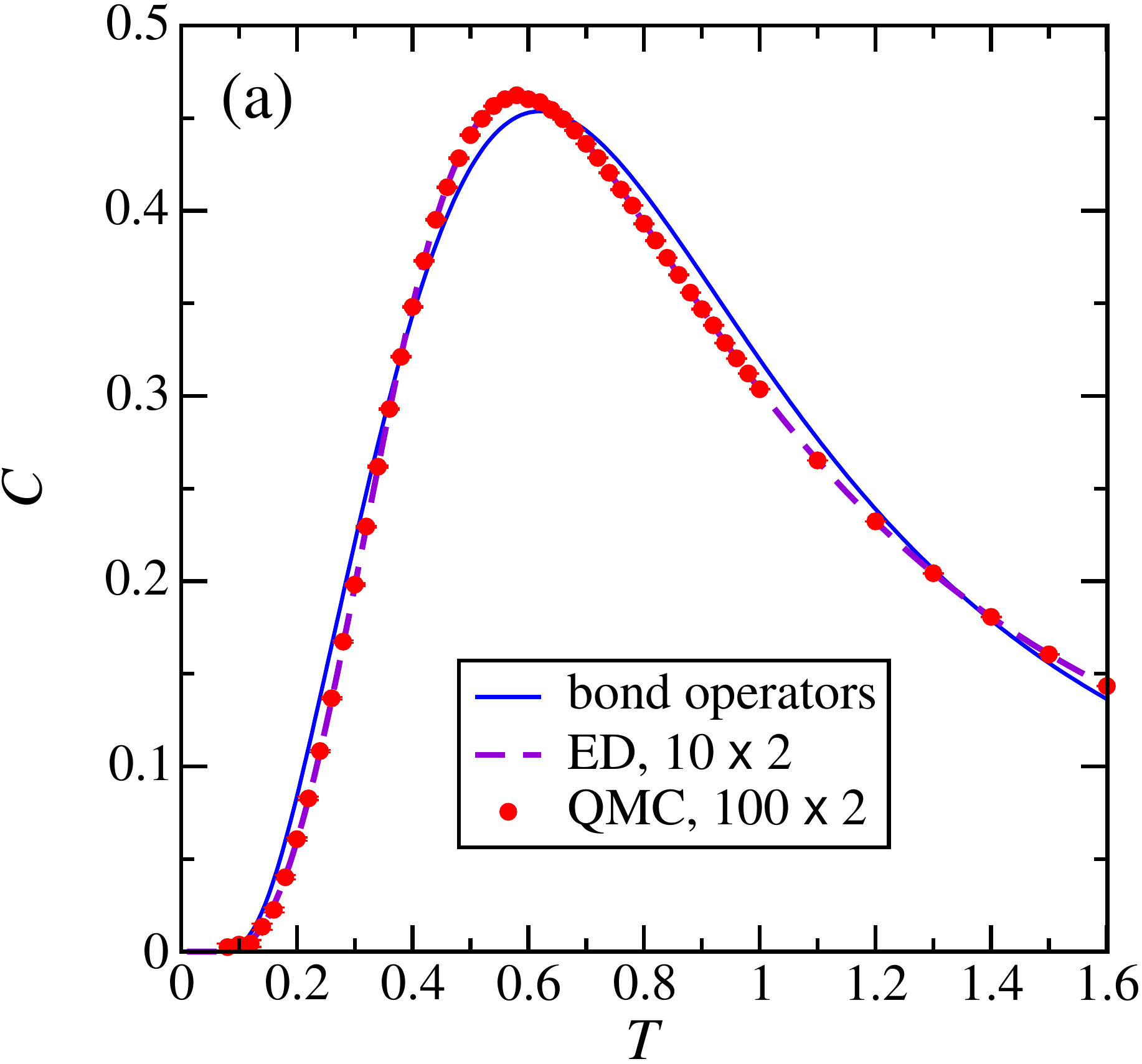}\hfill%
\includegraphics[width=0.49\columnwidth]{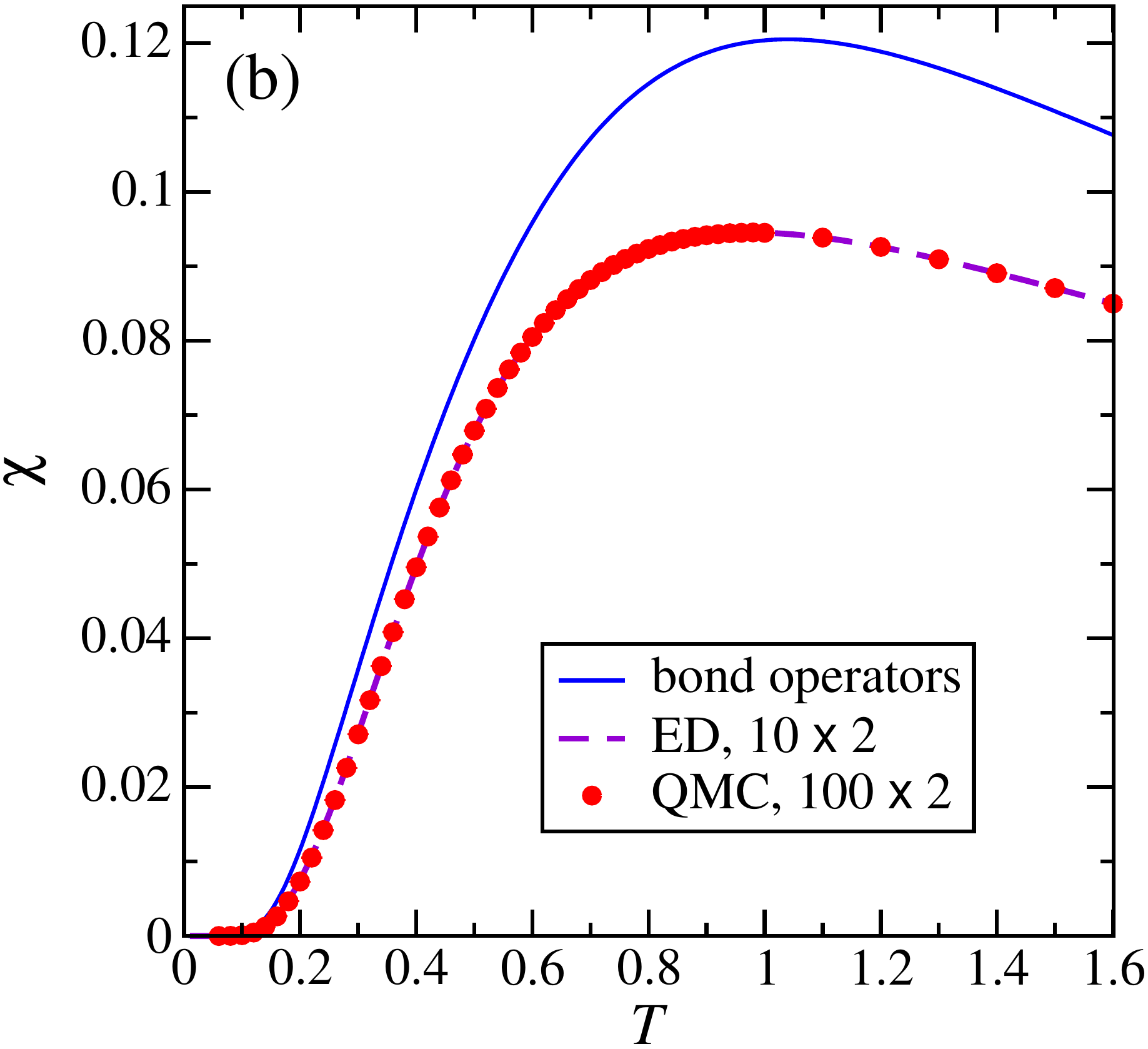}
\caption{Magnetic specific heat, $C$, and susceptibility, $\chi$, shown per 
spin for a partially frustrated ladder with rung coupling $J_\perp = 1.45$, 
leg coupling $J_\| = 1$, and diagonal coupling $J_\times = 0.2$. We compare 
QMC results obtained for ladders of $L = 100$ rungs with ED results for 
$L = 10$ and the bond-operator result for a corresponding unfrustrated ladder
with $J_\perp = 1.45$ and effective leg coupling $J_{\|,{\rm eff}} = 0.8$.}
\label{fig:thermoBreakJR1_45JD0_2}
\end{figure}

We turn next to the behavior of the frustrated ladder far 
from the phase transition. Remaining in the rung-singlet phase, 
Fig.~\ref{fig:thermoBreakJR1_45JD0_2} presents results for a point deep 
inside this regime, $J_\perp/J_\| = 1.45$, $J_\times/J_\| = 0.2$. Here we 
find that the data for $L = 10$ and $100$ rungs are indistinguishable, 
signaling a system with a short correlation length. For an analytical 
understanding of this case, we consider a bond-operator treatment 
\cite{rnr} of the frustrated ladder. In this approach, $J_\times$ appears 
on exactly the same footing as $J_\|$, giving an effective unfrustrated 
ladder with leg coupling $J_\| - J_\times$, which is expected to provide a 
good description of the thermodynamic response far from full frustration. 
In Fig.~\ref{fig:thermoBreakJR1_45JD0_2}, we illustrate this by comparing 
our QMC data with bond-operator results for a ladder with rung coupling 
$J_\perp = 1.45$ and leg coupling $J_{\|,{\rm eff}} = 0.8$, where the gap, 
$\Delta = 0.575 J_\perp$, corresponds to a correlation length $\xi \propto 
1/\Delta$ of order 3 lattice constants \cite{PhysRevLett.77.1865}. The 
effective unfrustrated model gives an excellent description of $C(T)$ for 
the weakly frustrated ladder, indicating that all the complexity of low-lying 
bound states is absent in this regime. However, we observe that the effective 
model is not able to reproduce in full the flattening of the peak in $\chi(T)$ 
caused by the presence of frustration. We comment that there is no longer any 
particularly low temperature scale characterizing the response of the system 
for these parameters.  

\subsection{Rung-triplet phase}

\label{sec:RungTriplet}

\begin{figure}[t!]
\includegraphics[width=0.49\columnwidth]{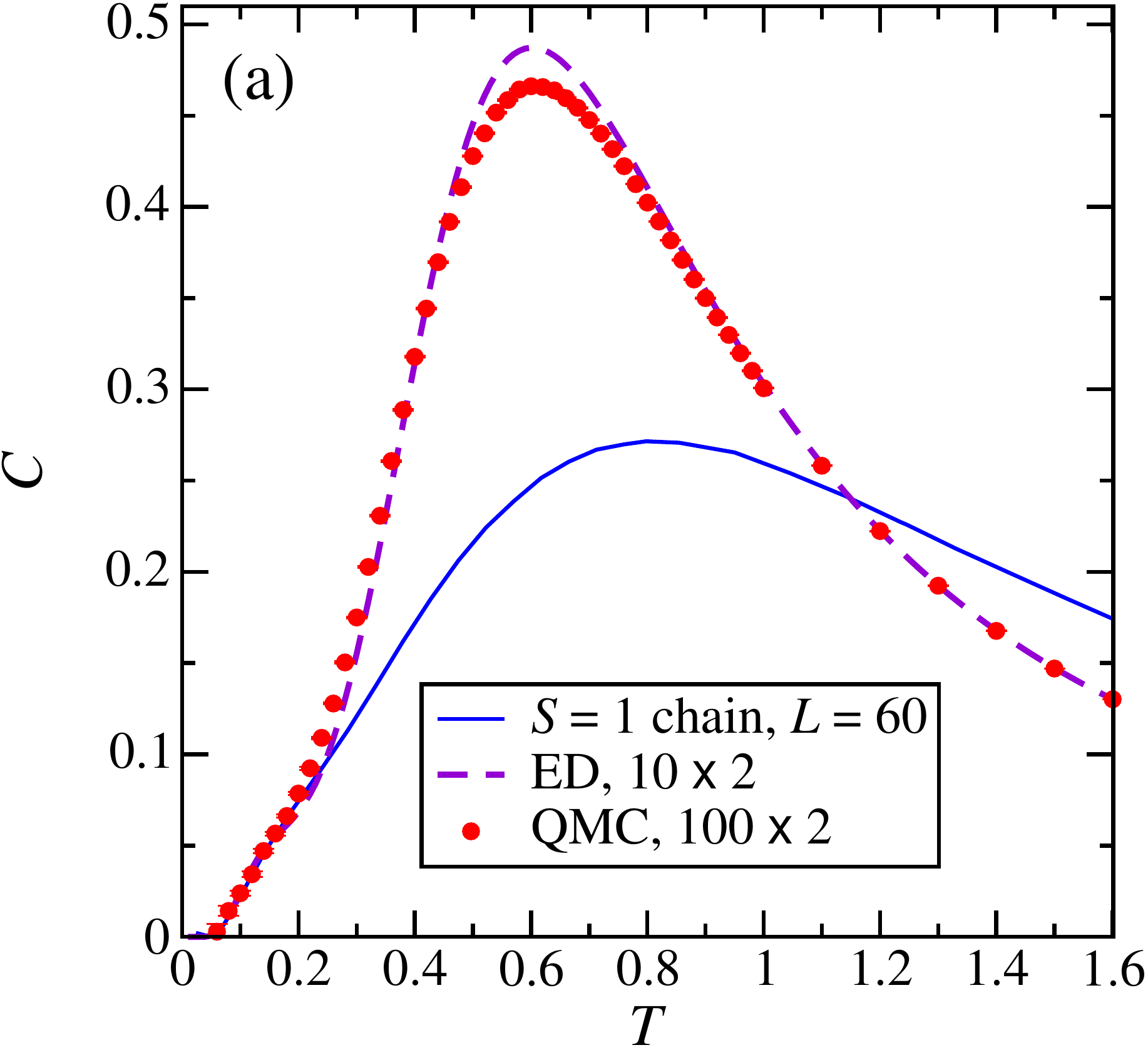}\hfill%
\includegraphics[width=0.49\columnwidth]{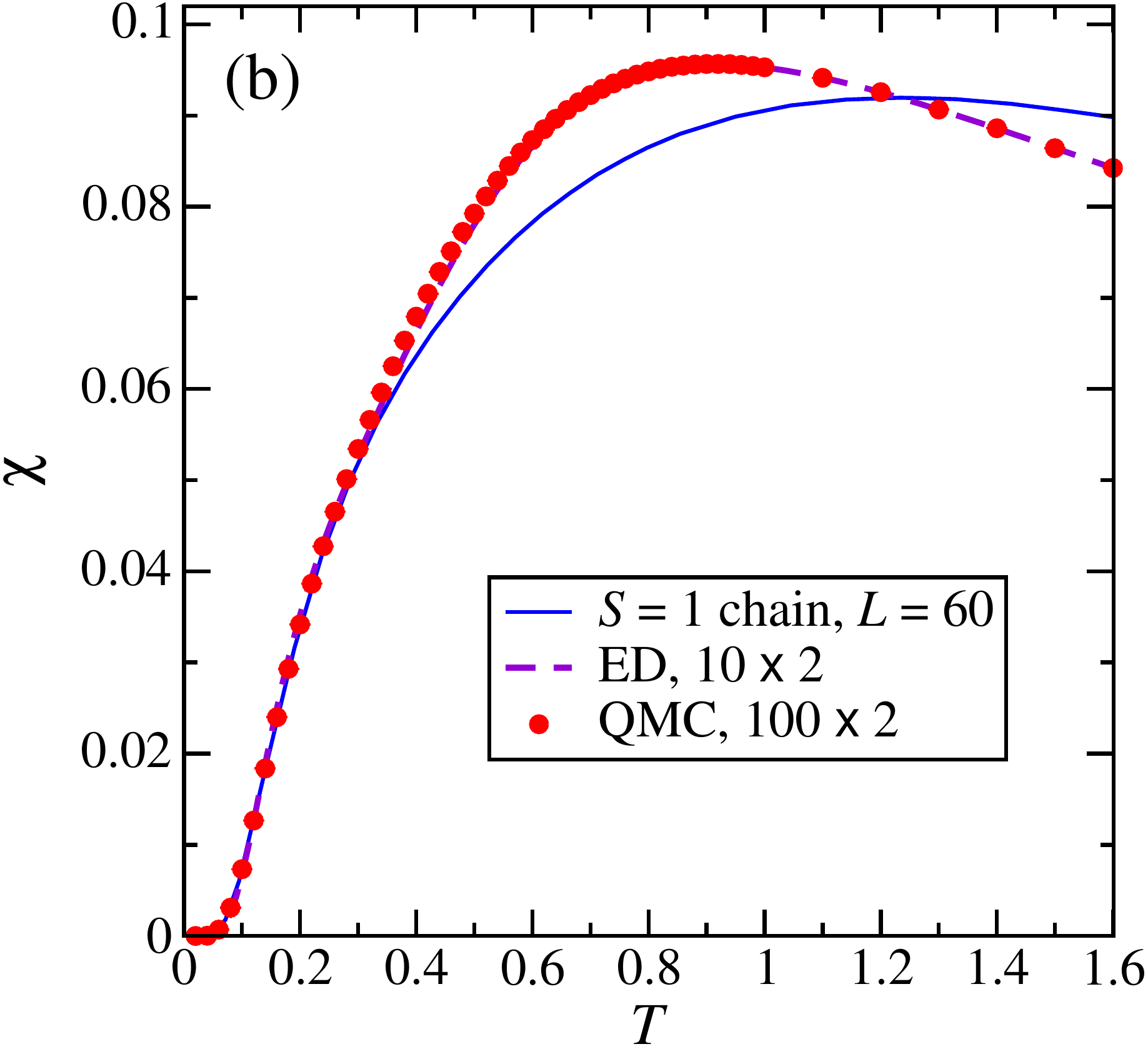}
\caption{Magnetic specific heat, $C$, and susceptibility, $\chi$, shown per 
spin for a partially frustrated ladder with rung coupling $J_\perp = 0.38$, 
leg coupling $J_\| = 1$, and diagonal coupling $J_\times = 0.9$. We compare 
QMC results obtained for ladders of $L = 100$ rungs with ED results for 
$L = 10$, and with QMC results for a spin-1 chain of $L = 60$ sites 
\cite{PhysRevB.79.104432} with $J_{\rm eff} = 0.95$, which are normalized 
to twice the number of spins in the chain.}
\label{fig:thermoBreakJR0_38JD0_9}
\end{figure}

Figure~\ref{fig:thermoBreakJR0_38JD0_9} presents the contrasting results for a 
point deep within the rung-triplet phase, $J_\perp/J_\| = 0.38$, $J_\times/J_\| = 
0.9$. From Eq.~(\ref{eq:exehGen}), the low-energy physics of this case should 
be similar to an $S = 1$ Heisenberg chain with an effective coupling constant 
$J_{\rm eff} = (J_\| + J_\times)/2 = 0.95\,J_\|$. We note first that finite-size 
effects in the specific heat [Fig.~\ref{fig:thermoBreakJR0_38JD0_9}(a)] 
are comparable to those observed in the fully frustrated ladder at $J_\times
 = J_\| = J_\perp = 1$ \cite{PhysRevB.93.054408}. These are to be expected 
because the ED system size, $L = 10$, is not significantly larger than the 
correlation length, $\xi \approx 6$, of the spin-1 chain at $T = 0$ 
\cite{PhysRevB.48.3844,PhysRevB.50.3037}. To test this comparison in full, 
Fig.~\ref{fig:thermoBreakJR0_38JD0_9} includes properly rescaled QMC results 
for a spin-1 Heisenberg chain with $L = 60$ sites \cite{PhysRevB.79.104432}. 
It is clear that these do match the low-temperature asymptotics of the 
frustrated ladder, and from them we identify the Haldane gap 
\cite{PhysRevB.48.3844,PhysRevB.50.3037}, $\Delta \approx 0.4105 \, J_{\rm eff} 
\approx 0.39\,J_\|$, as the lowest energy and temperature scale for the 
parameters of Fig.~\ref{fig:thermoBreakJR0_38JD0_9}. At higher temperatures,
we observe strong additional contributions beyond the effective spin-1 chain, 
in particular in $C(T)$, and these can be attributed to the thermal population 
of mostly localized rung-singlet excitations, as discussed for the fully 
frustrated ladder in Ref.~\cite{PhysRevB.93.054408}. 

\subsection{Phase transition and weakly coupled chains}

\label{sec:Decoupl}

We conclude our survey of the phase diagram by returning to its most complex 
region, the phase-transition line. Here we will show that it remains possible 
to obtain highly accurate results by QMC in the rung basis, whereas ED is
affected by severe finite-size effects at low temperatures.

\begin{figure}[t!]
\includegraphics[width=0.49\columnwidth]{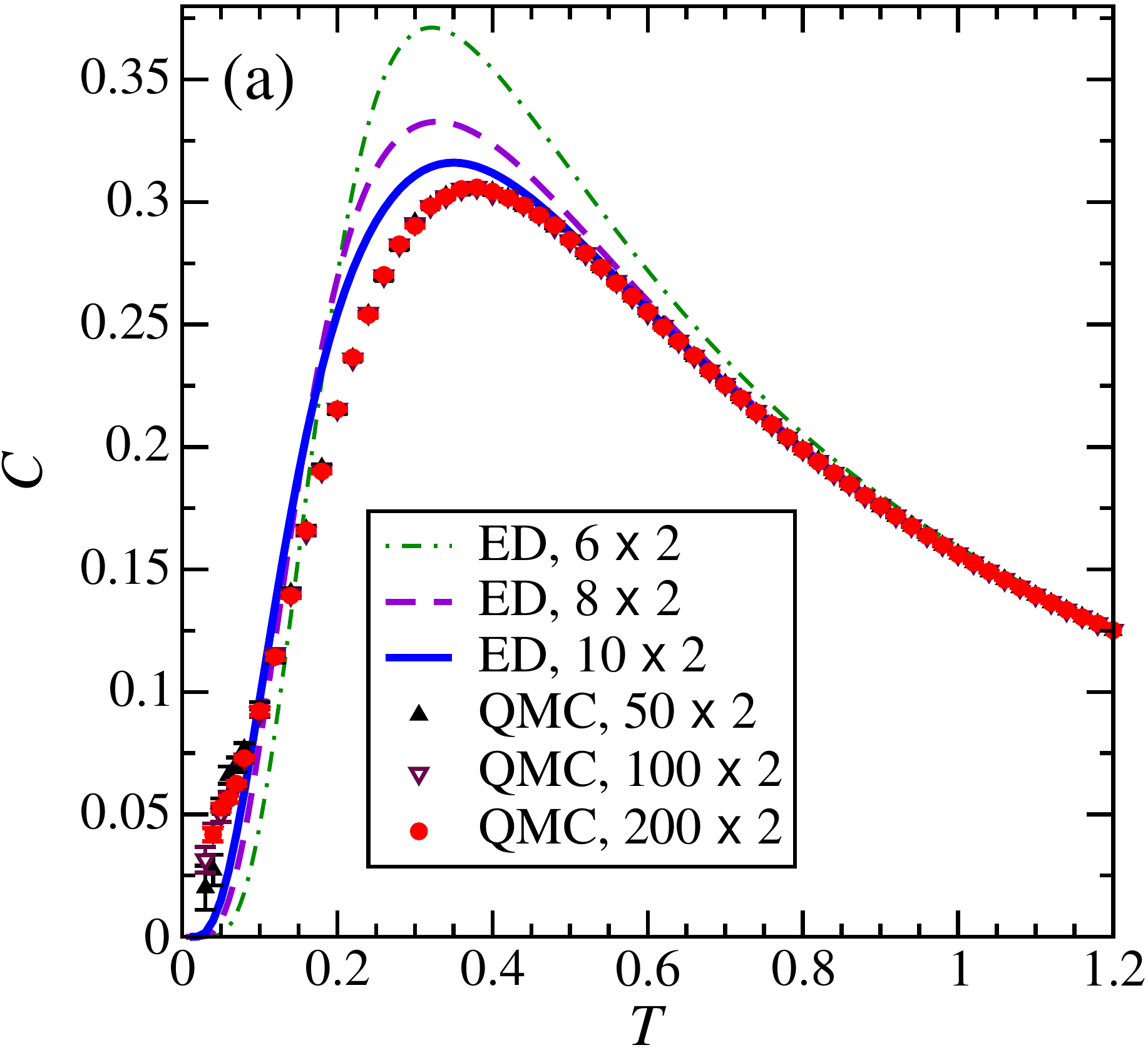}\hfill%
\includegraphics[width=0.49\columnwidth]{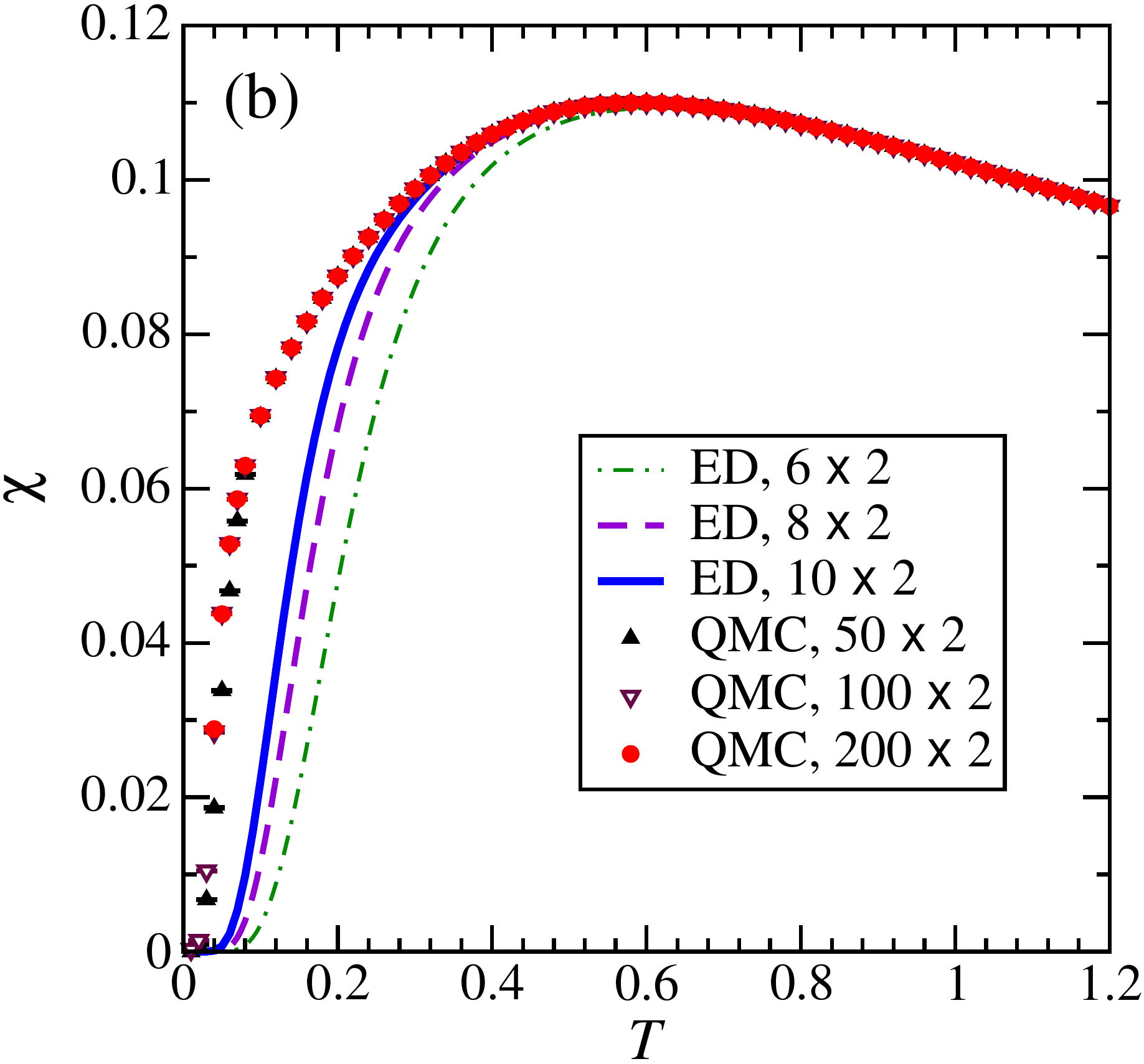}
\caption{Magnetic specific heat, $C$, and susceptibility, $\chi$, shown per 
spin for a ladder with rung coupling $J_\perp = 1$, leg coupling $J_\| = 1$, 
and diagonal coupling $J_\times = 0.599$. We compare QMC results obtained for 
ladders of $L = 50$, 100, and 200 rungs with ED results for $L = 6$, 8, and 
10.} 
\label{fig:thermoBreakJR1JD0_599}
\end{figure}

Figure \ref{fig:thermoBreakJR1JD0_599} shows thermodynamic results for the 
point $J_\perp/J_\| =  1$, $J_\times/J_\| =  0.599$, which is located on the 
transition line, but moved from Figs.~\ref{fig:thermoFFL1_4} and 
\ref{fig:thermoBreakJR1_32JD0_9} in the direction of weakly coupled chains. 
Once again, all QMC results with $L \ge 50$ rungs are consistent and thus 
can be considered as representing the thermodynamic limit. However, the 
finite-size effects visible in the ED results are distinctly enhanced, by 
comparison with Fig.~\ref{fig:thermoBreakJR1_32JD0_9}, at all temperatures 
below the peaks in $C(T)$ and $\chi(T)$. A $T = 0$ density-matrix 
renormalization-group (DMRG) investigation \cite{Wang2000} found a 
gap $\Delta/J_\| \approx 0.13$ for the parameters of 
Fig.~\ref{fig:thermoBreakJR1JD0_599}. Despite the appearance of anomalous 
features such as the shoulder in the low-temperature specific heat, visible 
in Fig.~\ref{fig:thermoBreakJR1JD0_599}(a), our results are indeed consistent 
with a finite gap, albeit one with the nature of an emergent low energy scale. 
As a consequence, we conclude that the $T = 0$ transition from the rung-singlet 
to the rung-triplet phase remains of first-order type for $J_\times/J_\| \ge 0.6$.

\begin{figure}[t!]
\includegraphics[width=0.49\columnwidth]{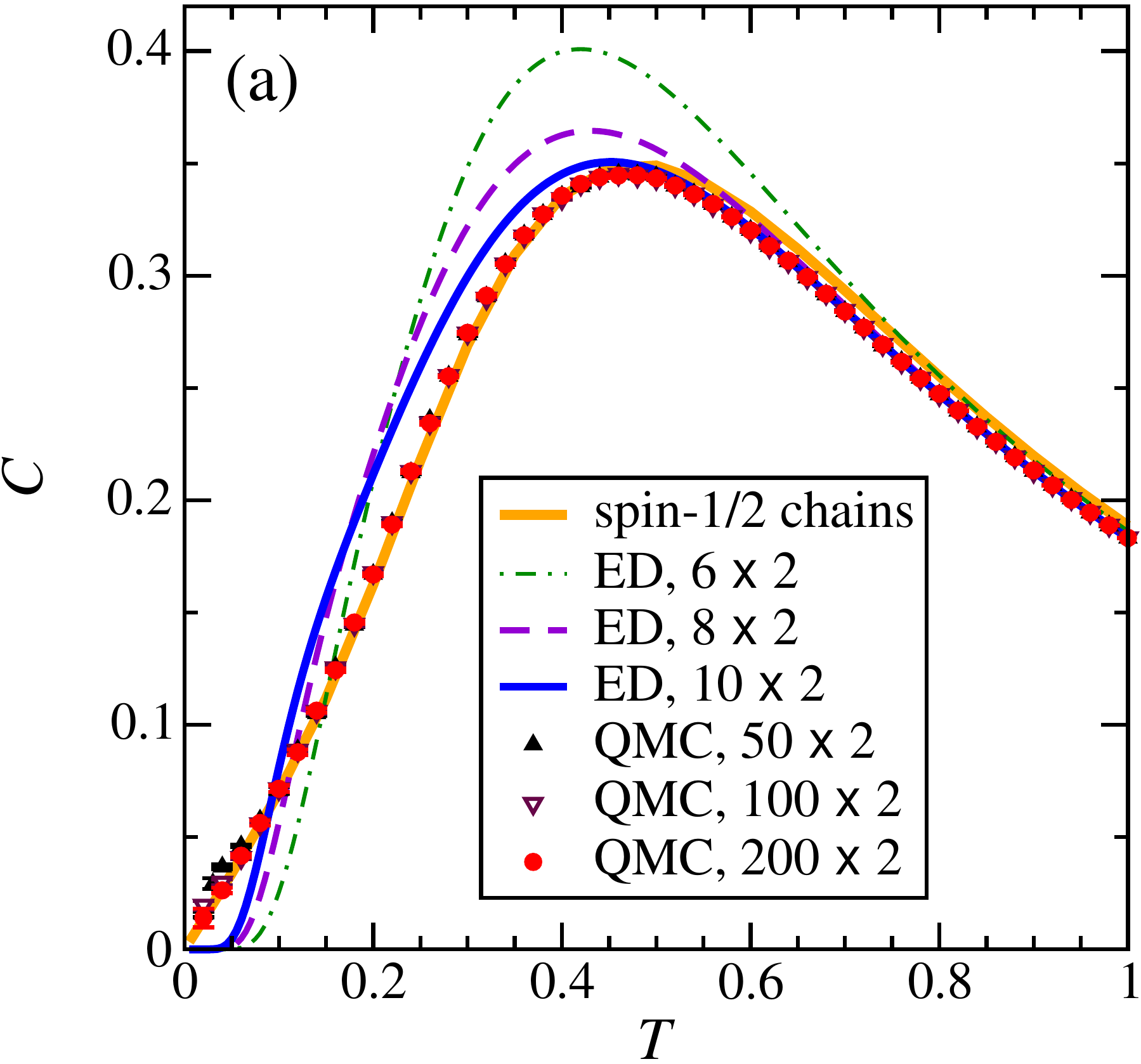}\hfill%
\includegraphics[width=0.49\columnwidth]{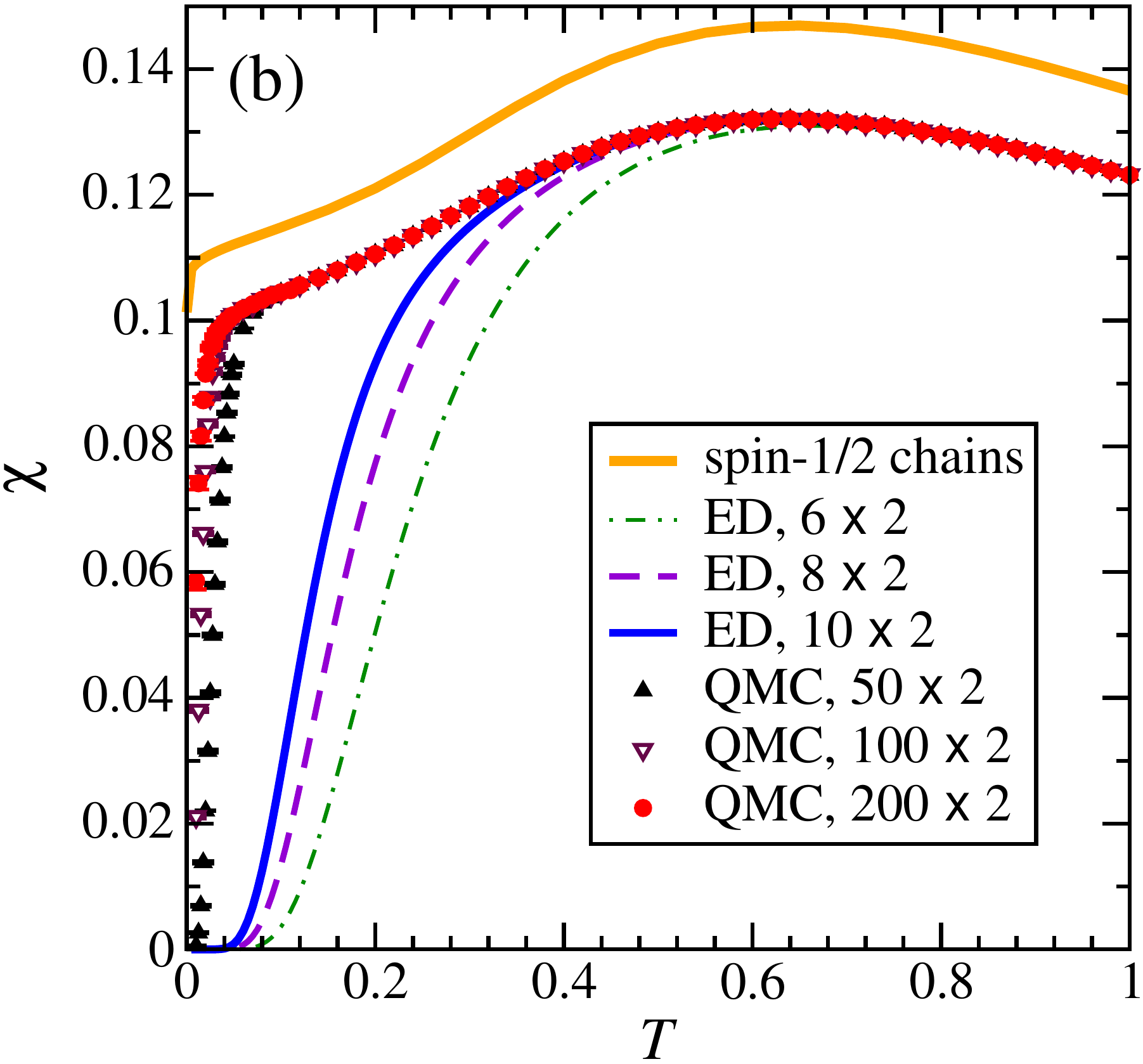}
\caption{Magnetic specific heat, $C$, and susceptibility, $\chi$, shown per 
spin for a ladder with rung coupling $J_\perp = 0.38$, leg coupling $J_\| = 1$,
and diagonal coupling $J_\times = 0.196$. We compare QMC results obtained for 
ladders of $L = 50$, 100, and 200 rungs with ED results for $L = 6$, 8, and 10, 
as well as with exact results for two decoupled spin-1/2 chains, with coupling 
constant $J_\| = 1$, in the thermodynamic limit \cite{PhysRevLett.84.4701}.}
\label{fig:thermoBreakJR0_38JD0_196}
\end{figure}

Moving yet further in the direction of weak interchain coupling, 
Fig.~\ref{fig:thermoBreakJR0_38JD0_196} shows results for a ladder 
with $J_\perp/J_\| =  0.38$ and $J_\times/J_\| =  0.196$. The average 
sign appearing in the QMC simulations for this system was shown in 
Fig.~\ref{fig:signJR0_38JD0_196} and we observe that its deviations from 
unity remain sufficiently small that they do not impact the accuracy of the
QMC results. However, the finite-size effects in the ED calculations are now 
enhanced very dramatically, and even in the QMC simulations it is clear that 
the $L = 200$ data are required to ensure a good approximation to the 
thermodynamic limit for the features of $C$ and $\chi$ at the lowest 
temperatures; in particular, the susceptibility is inordinately sensitive 
at $T/J_\| < 0.05$ [Fig.~\ref{fig:thermoBreakJR0_38JD0_196}(b)]. From our 
observation (Sec.~\ref{sec:SignNum}) that the sign problem arises due to a 
boundary term, it is naturally sensitive to the correlation length of the 
system and thus its most serious manifestations (Figs.~\ref{fig:signScan} 
and \ref{fig:signJR0_38JD0_196}), as well as the most serious finite-size 
effects (Fig.~\ref{fig:thermoBreakJR0_38JD0_196}), occur when the correlation 
length is largest, explaining directly why $1 - \langle \text{sign} \rangle$ 
at fixed temperature acts as an excellent indicator of the phase transition 
in parameter space. 

To interpret the results in Fig.~\ref{fig:thermoBreakJR0_38JD0_196}, we 
show the exact specific heat \cite{PhysRevLett.84.4701} and susceptibility 
\cite{PhysRevLett.73.332,PhysRevLett.84.4701} of two isolated spin-1/2 chains.
Our numerical data for $C(T)$ converge to a curve whose low-temperature 
behavior is captured remarkably well by a pair of decoupled chains, implying 
that for this parameter regime the rung and diagonal interactions act to 
cancel each other. The shape of $\chi(T)$ for this ladder also follows 
very closely the result for two decoupled chains, where it is known to 
approach a logarithmic singularity at low $T$, although the absolute scale 
is renormalized by a factor of approximately $0.9$. We expect that this 
deviation reflects in part the sensitivity of the $S = 1/2$ degrees of freedom 
in the ladder legs to the confining effects of the relevant rung-coupling 
perturbation and in part the sensitivity to this coupling of matrix-element 
effects. Thus our results quantify the extent to which the ladder legs may 
be regarded as effectively decoupled in the presence of finite but mutually 
antagonistic rung and diagonal interactions. An essential qualitative 
observation is that our numerical data for $C$ and $\chi$ are most easily 
reconciled with a direct and continuous transition at $T = 0$ between the 
rung-singlet and rung-triplet phases at $J_\perp = 0.38$, with no evidence 
for the presence of an intermediate phase.

\subsection{Frustrated ladder physics}

\label{sec:psum}

We conclude our discussion of the physics of the frustrated two-leg ladder 
with a brief summary. With the possible exception of a vanishingly narrow 
region at small interchain coupling, the frustrated ladder has only two 
ground states, a rung-singlet phase when the rung coupling significantly 
exceeds the leg and diagonal couplings and a rung-triplet phase in the opposite 
situation. Far from the quantum phase transition separating the two states, 
the rung-singlet phase can be described rather well by an effective 
unfrustrated ladder model (Fig.~\ref{fig:thermoBreakJR1_45JD0_2}) while 
the rung-triplet phase contains the physics of both quasi-localized 
bound states and the extended states of an effective spin-1 chain 
(Fig.~\ref{fig:thermoBreakJR0_38JD0_9}). 

In the strongly coupled and strongly frustrated ladder, the transition is 
strongly first-order. This regime is characterized by large numbers of 
low-lying bound states, which manifest themselves in the sharp peak in $C(T)$ 
and very low onset temperature in $\chi(T)$ (Figs.~\ref{fig:thermoFFL1_45} and 
\ref{fig:thermoBreakJR1_4JD0_9}). However, exactly at the transition, the 
diverging number of these states causes the sharp specific-heat peak to 
vanish (Figs.~\ref{fig:thermoFFL1_4} and \ref{fig:thermoBreakJR1_32JD0_9}). 
It is clear that the behavior of the manifold of low-lying bound states 
around the transition is not very sensitive to whether the ladder is perfectly 
frustrated (Figs.~\ref{fig:thermoFFL1_4} and \ref{fig:thermoBreakJR1_32JD0_9}) 
or not (Figs.~\ref{fig:thermoFFL1_45} and \ref{fig:thermoBreakJR1_4JD0_9}).

As the phase-transition line is followed to weaker interchain couplings, there 
is a clear change of behavior, consistent with a closing of the gaps of both 
the rung-singlet and -triplet phases (Fig.~\ref{fig:thermoBreakJR1JD0_599}). 
Previous DMRG and field-theoretical studies indeed suggest a gapless state  
and continuous transitions across the line in this regime. While our QMC 
results cannot exclude the possibility that this topological transition 
remains of first order with an exponentially small gap, it is clear that 
the physics of the frustrated ladder with weakly coupled chains 
(Fig.~\ref{fig:thermoBreakJR0_38JD0_196}) is very different from 
the strongly coupled regime 
(Figs.~\ref{fig:thermoFFL1_45}--\ref{fig:thermoBreakJR1_32JD0_9}).
We suggest that DMRG may be the most appropriate technique for revisiting
the nature of the quantum phase transition on the approach 
to two decoupled chains, for example by computing 
the central charge and the evolution of the gaps in both phases. 

\section{Summary and perspectives}

\label{sec:Sum}

We have demonstrated that the sign problem which plagues QMC studies of 
frustrated quantum spin models can be rendered so weak as to be irrelevant 
in certain classes of system. Taking the example of the frustrated two-leg 
spin-1/2 ladder, we have shown that efficient QMC simulations can be performed
throughout the entire phase diagram, and that ladders of size 200 rungs access 
the thermodynamic limit in all cases. The key to the success of our improved 
QMC approach is to rewrite the Hamiltonian in a different basis, which can be 
regarded as forming a composite-spin system out of the original single-site 
basis; in the ladder, the natural choice is a basis of ladder rungs. 

Complete elimination of the sign problem is achieved whenever a rewriting 
of the form of Eq.~(\ref{eq:exeh}) is possible, which leads to a composite-spin
model whose basis units themselves form an unfrustrated lattice 
\cite{PhysRevB.93.054408,PhysRevLett.117.197203}. This situation arises not 
only in the fully frustrated case of the present ladder model but also in a 
considerable number of highly frustrated spin models whose frustration 
results in local conservation laws \cite{Honecker2000,PhysRevB.92.115111,
LTP07,azurite11,PhysRevB.86.054412,NUZ97,KKS96,NUZ98,RIS98,PhysRevB.59.13806,
PhysRevB.62.5558,PhysRevLett.84.1808,PhysRevB.65.054420,PhysRevB.66.184402,
CB02,PhysRevB.67.174401,PhysRevB.72.024443,PhysRevB.74.144430,
PhysRevB.82.214412,PhysRevB.93.144413,PhysRevB.93.235150,KBDSR17}
for clusters forming a bipartite lattice. While a square-lattice bilayer 
analog of the fully frustrated ladder has recently been studied by QMC 
simulations \cite{PhysRevLett.117.197203,NgYang2017}, further cases in the 
same class remain to be investigated. Although the example of the spin-1/2 
$J_1$--$J_2$ chain \cite{PhysRevB.57.R3197,PhysRevB.58.2411} and some of our 
own results for ladders far from perfect frustration demonstrate that a 
rewriting of the Hamiltonian is not a prerequisite for performing QMC, it 
remains true that appropriately chosen cluster bases provide a general tool 
for optimizing the efficiency of QMC simulations.

The fact that the sign problem remains mild in the entire space of parameters
for the ladder with $J_\times \ne J_\|$ can be traced to the fact that only 
$\vec{D} \cdot \vec{D}$ terms appear in the rung-basis Hamiltonian of 
Eq.~(\ref{eq:exehGen}). We believe that such favorable behavior is a 
generic feature of models where a global exchange symmetry of the two ends 
of each dimer ensures the absence of $\vec{T} \cdot \vec{D}$ terms. This 
includes frustrated magnets in higher dimensions, where few accurate 
numerical methods exist. However, $\vec{T} \cdot \vec{D}$ terms do appear 
in more general models when expressed in a dimer basis, not least the 
asymmetrically frustrated ladder, and the Shastry-Sutherland model 
\cite{SRIRAMSHASTRY19811069,Miyahara2003} provides a key example in two 
dimensions. We defer to future studies a detailed investigation of the 
extent to which suitably chosen cluster bases render QMC simulations 
feasible for higher-dimensional models. Here we comment only that the 
search for such an optimal cluster basis in any given frustrated spin 
system may also be extended by performing a systematic scan of the 
manifold of basis transformations \cite{PhysRevB.92.195126}.
 
We remark also that the sign problem remaining for the frustrated ladder 
in the rung basis takes a new and quite unconventional form, in which it 
becomes weaker as a function of increasing system size and is maximal at a 
finite temperature before the average sign recovers to unity at low $T$. 
From a practical standpoint, these results offer the possibility of working 
around the sign problem for any given parameter set. From an analytical 
point of view, they provide new insight into the nature and number of the 
negative-weight configurations, which in the frustrated ladder model appear 
to be a boundary problem only (and hence a set of vanishing measure in the 
thermodynamic limit). Finally, once QMC simulations have been demonstrated 
to work efficiently for the computation of static thermodynamic properties, 
it will be of great value to apply them also to the computation of dynamical 
response functions at finite temperatures.

\section*{Acknowledgements}

We thank F.~Alet, K.~Damle, R.~Noack, G.~Radtke, and O.~Vaccarelli for helpful 
discussions.

\paragraph{Funding information}

This work was supported by the DFG research unit FOR1807 and by the Swiss NSF.

\bibliography{fladder-detuned}

\end{document}